\documentclass{aastex}

\shorttitle{Optical Monitoring of Mrk 501}
\shortauthors{Xiong et
al.}

\begin{document}
\title{MULTI-COLOR OPTICAL MONITORING OF MRK 501 FROM 2010 TO 2015}

\author{Dingrong Xiong,}
\affil{National Astronomical Observatories/Yunnan Observatories,
Chinese Academy of Sciences, Kunming 650011, China\\
University of Chinese Academy of Sciences, Beijing 100049, China}

\author{Xiong Zhang,   Tingfeng Yi,}
\affil{Department of Physics, Yunnan Normal University, Kunming
650500, China}

\author{Jinming Bai,}
\affil{National Astronomical Observatories/Yunnan Observatories,
Chinese Academy of Sciences, Kunming 650011, China}

\author{Fang Wang,}
\affil{National Astronomical Observatories/Yunnan Observatories,
Chinese Academy of Sciences, Kunming 650011, China\\
University of Chinese Academy of Sciences, Beijing 100049, China}

\author{Hongtao Liu,}
\affil{National Astronomical Observatories/Yunnan Observatories,
Chinese Academy of Sciences, Kunming 650011, China}

\author{Yonggang Zheng and Haojing Zhang}
\affil{Department of Physics, Yunnan Normal University, Kunming
650500, China} \email{kmzhanghj@163.com}

\begin{abstract}

We have monitored the BL Lac object Mrk 501 in optical $V$, $R$ and
$I$ bands from 2010 to 2015. For Mrk 501, the presence of strong
host galaxy component can affect the results of photometry. After
subtracting the host galaxy contributions, the source shows intraday
and long-term variabilities for optical flux and color indices. The
average variability amplitudes of $V$, $R$ and $I$ bands are
$22.05\%, 22.25\%, 23.82\%$ respectively, and the value of duty
cycle 14.87 per cent. A minimal variability timescale of 106 minutes
is detected. No significant time lag between $V$ and $I$ bands is
found on one night. The bluer-when-brighter (BWB) trend is dominant
for Mrk 501 on intermediate, short and intraday timescales which
supports the shock-in-jet model. For the long timescale, Mrk 501 in
different state can have different BWB trend. The corresponding
results of non-correcting host galaxy contributions are also
presented.

\end{abstract}


\section{INTRODUCTION}

Blazars are the most extreme subclass of active galactic nuclei
(AGNs), whose jets point in the direction of the observer, and they
are characterized by large amplitude and rapid variability at all
wavelengths, high and variable polarization, superluminal jet speeds
and compact radio emission (Angel \& Stockman 1980; Urry \& Padovani
1995). Blazars are often divided into two subclasses of BL Lacertae
objects (BL Lacs) and flat spectrum radio quasars (FSRQs). FSRQs
have strong emission lines, while BL Lac objects have only very weak
or non-existent emission lines. The classic division between FSRQs
and BL Lacs is mainly based on the equivalent width (EW) of the
emission lines. Objects with rest frame EW$>5$ {\AA} are classified
as FSRQs (e.g. Urry \& Padovani 1995; Scarpa \& Falomo 1997). The
broad-band spectral energy distribution (SED) of blazars are usually
bimodal. The lower bump peak is in the IR-optical-UV band and the
higher bump in the GeV-TeV gamma-ray band (Ghisellini et al. 1998;
Abdo et al. 2010). Abdo et al. (2010) and Ackermann et al. (2011)
proposed the synchrotron peak frequency
$\nu^s_{\rm{peak}}<10^{14}{\rm Hz}$ for low-synchrotron-peaked
blazar (LSP), $10^{14}{\rm Hz}<\nu^s_{\rm{peak}}<10^{15}{\rm Hz}$
for intermediate-synchrotron-peaked blazar (ISP), $10^{15}{\rm
Hz}<\nu^s_{\rm{peak}}$ for high-synchrotron-peaked blazar (HSP).

Blazars show variability on different timescales from years down to
minutes (e.g. Poon et al. 2009). Blazar variability can be broadly
divided into intraday variability (IDV) or micro-variability,
short-term variability (STV) and long-term variability (LTV).
Brightness changes of a few tenth of a magnitude in a time scale of
 tens of minutes to a few hours is often known as IDV (Wagner \& Witzel
 1995). STV have time scales of days to weeks, even months and LTV ranges
 from months to years (Gupta et al. 2008; Dai et al. 2015).
 Understanding blazar variability is one of the major issues of
 AGNs. Variability can shed light on the location, size, structure
 and dynamics of the emitting regions and radiation mechanism
 (Ciprini et al. 2003, 2007; Dai et al. 2015). Optical variability
 are often associated with color/spectral behavior in blazars which
 can be used to explore the emission mechanism (e.g. Gu \& Ai 2011).

 The BL Lac object Mrk 501 (R.A.=$16^{\rm h}45^{\rm m}52^{\rm s}.22$, decl.=
 $39^{\rm \circ}45^{\rm '}36^{\rm ''}.6$, J2000, redshift $z=0.034$, HSP)
 is one of the brightest extragalactic sources
 in the X-ray/TeV sky, and the second extragalactic object
 identified as a very high energy $\gamma$-ray emitter (Abdo et al.
 2011). The flux and spectral variation of Mrk 501 have been
 extensively studied over the entire electromagnetic spectrum (e.g. Stickel et al. 1993; Heidt \& Wagner 1996;
 Quinn et al. 1996; Catanese et al. 1997;
  Pian et al. 1998; Aharonian et al. 1999a, 1999b; Ghosh et al. 2000; Xie et al. 1999, 2001; Xue \& Cui 2005;
 Albert et al. 2007; Gupta et al. 2008; Rodig et al. 2009;
 Abdo et al. 2011; Neronov et al. 2012; Bartoli et al. 2015; Wierzcholska et al. 2015;
Shukla et al. 2015; Aleksic et al. 2015). In 1997, Mrk 501 went into
a state with surprisingly high activity and strong variability and
became more than a factor of 10 brighter (above 1 TeV) than the Crab
Nebula (Aharonian et al. 1999a, 1999b). In 1998 - 1999, the mean VHE
$\gamma$-ray flux dropped by an order of magnitude, and the overall
VHE spectrum softened significantly (Aharonian et al. 2001). The
fastest $\gamma$-ray flux variability on a timescale of minutes was
observed in 2005 in the VHE band (Albert et al. 2007). The
significant ``harder when brighter'' spectral variability was also
detected. Compared with low-activity state, the ``harder when
brighter'' behavior are more pronounced at high-activity level
(Anderhub et al. 2009; Acciari et al. 2011). In 2009, Abdo et al.
(2011) observed relatively mild flux variations with the Fermi-LAT,
and detected remarkable spectral variability. But these spectral
changes do not correlate with the measured flux variations above 0.3
GeV. Ikejiri et al. (2011) performed monitoring of 42 blazars in the
optical and near-infrared bands from 2008 to 2010, and found that
Mrk 501 exhibits the ``bluer-when-brighter'' trend. Wierzcholska et
al. (2015) presented the results of a long-term optical monitoring
of blazars, and found a significant bluer-when-brighter behavior for
Mrk 501. In the optical region, the presence of a strong host galaxy
component can affect measured aperture photometry at least in two
way: (i) the host galaxy adds flux to the measurement aperture; (ii)
in some cases the host galaxy contribution can exceed the nuclear
flux by a large margin (Nilsson et al. 2007). The results of Nilsson
et al. (2007) also show a prominent host galaxy component for Mrk
501. So Mrk 501 need a correction for the host galaxy contribution
in the optical band. However, From previous studying about the
optical flux and spectral variability of Mrk 501, we can find that
most of authors neglected the host galaxy contributions.

Mrk 501 is a very important source for studying GeV/TeV emission.
The SED of Mrk 501 has two bumps occurring at keV and GeV/TeV
energies. Generally, The first bumps is caused by synchrotron
radiation from the electron, but the origin of the GeV/TeV hump is
unclear. Although Mrk 501 has been observed over the last two
decades in the entire electromagnetic spectrum, the existing
multi-wavelength data could not provide an explicit answer for the
physical mechanisms that are responsible for the production of the
GeV/TeV hump (e.g. Shukla et al. 2015). The SED and the
multi-wavelength correlations of Mrk 501 have been intensively
studied in the past, but the nature of this object is still far from
being understood. The main reason is lack of simultaneous
multi-wavelength data during long periods of time (Abdo et al.
2011). Since the launch of the Fermi satellite, we have entered in a
new era of blazars research (Abdo et al. 2009). The Fermi LAT with
the higher sensitivity and larger energy range has become a crucial
tool for studying Mrk 501. However, it is important to emphasize
that blazars can vary their emitted power by one or two orders of
magnitude on short timescales, and that they emit radiation over the
entire observable electromagnetic spectrum. Therefore the
information from Fermi-LAT alone is not enough to understand the
broadband emission of Mrk 501, and simultaneous data in other
frequency ranges are required. Then long-term optical studies are
also so helpful for understanding the nature of Mrk 501.

In view of these facts, we monitored the source from 2010 to 2015.
After correcting the host galaxy contribution, we analyze the
variability and spectral properties of Mrk 501. This paper is
organized as follows. The observations and data analysis is
described in Section 2. Section 3 presents the results. Discussion
and conclusions are reported in Section 4. A summary is given in
Section 5.

\section{OBSERVATIONS AND DATA ANALYSIS}

Our optical monitoring program of Mrk 501 was carried out with the
1.02 m optical telescope located at the the Yunnan Astronomical
Observatory (YAO) of China. The field of view of the CCD image is
7.3 $\times$ 7.3 arcmin$^2$ and its pixel scale is 0.21
arcsec/pixel. The telescope was equipped with an Andor DW436 CCD
(2048 $\times$ 2048 pixels) camera at the Cassergrain focus
($f=13.3$~m), with pixel size 13.5 $\times$ 13.5 $\mu m^2$. The
readout noise and gain were 6.33 electrons and 2.0 electrons/ADU,
respectively, with 2 $\mu$s (readout time per pixel) or 2.29
electrons and 1.4 electrons/ADU with 16 $\mu$s (Dai et al. 2015).
The standard Johnson broadband filters were used. The typical
integration time were 150 - 300 s for the $I$ and $R$ filters and
400 s for the $V$ filter. The optical observations in the $V$, $R$,
and $I$ bands are in a cyclic mode. Each cycle of $V$, $R$, and $I$
bands photometry was usually completed within 10-20 minutes, i.e.
quasi-simultaneously. Flat-field images were taken at dusk or dawn
when possible. The bias images were taken at the beginning and at
the end of the observations, while the dark images were taken at the
end of the observations. After correcting flat-field, bias and dark,
aperture photometry was performed using the APPHOT task of
IRAF\footnote{\small{IRAF is distributed by the National Optical
Astronomy Observatories, which are operated by the Association of
Universities for Research in Astronomy, Inc., under cooperative
agreement with the National Science Foundation.}}. The aperture
radius of 2$\times$FWHM or 1.5$\times$FWHM was selected, considering
the best S/N ratio. Following Zhang et al. (2004, 2008), Fan et al.
(2014), Bai et al. (1998), we can obtain the blazar magnitude from
two standard star in the same frame ($\frac{m_1+m_2}{2}$, $m_1$ is
the blazar magnitude obtained from standard star 1 and $m_2$ from
standard star 2). We observed 6 comparison stars on the same field.
Star 1, which is the brightest comparison star, is used as the first
standard star. The second standard star is chosen from the two
limits: its brightness is comparable with the source or a little
fainter than the source; there is the smallest variation in
differential magnitudes between standard star 1 and standard star 2.
The finding chart of Mrk 501 is obtained from the
webpage\footnote{http://www.lsw.uni-heidelberg.de/projects/extragalactic/charts/1652+398.html}.
The magnitudes of comparison stars in the field of Mrk 501 are taken
from Villata et al. (1998) and Fiorucci \& Tosti (1996). The rms
errors of the photometry on a certain night are calculated from the
two comparison stars, star $S_1$ and star $S_2$, in the usual way:
\begin{equation}
\sigma=\sqrt{\sum \frac{m_i-\overline{m}}{N-1}},~~~~i=1,2,3,...,N,
\end{equation}
where $m_i = (m_{\rm S_1}-m_{\rm S_2})_{i}$ is the differential
magnitude of stars $S_1$ and $S_2$, while
$\overline{m}=\overline{m_{S_1}-m_{S_2}}$ is the averaged
differential magnitude over the entire data set, and $N$ is the
number of the observations on a given night. The variability
amplitude (Amp) can be calculated by (Heidt \& Wagner 1996)
\begin{equation}
{\rm Amp}=100\times \sqrt{(A_{max}-A_{min})^2-2\sigma^2}~{\rm per
cent},
\end{equation}
where $A_{max}$ and $A_{min}$ are the maximum and minimum magnitude,
respectively, of  the light curve for the night being considered;
and $\sigma$ is the same as what is described above.

Our monitoring program for Mrk 501 was divided into five periods.
The first period was from 2010 April 1-3, the second from 2012 April
22-25, the third from 2013 April 1-4, the fourth from 2014 April 17
- May 8, and the fifth from 2015 April 6 - 2015 May 25. Excluding
the nights with bad weather and those devoted to other targets, the
actual number of observations for Mrk 501 is 29 nights. The
observation log is given in Table 1 where we have listed observation
date, time spans, time resolutions and number of data points for
each date in different band. The results of observations are given
in Table 2-4 for filters $I$, $R$ and $V$.

\section{RESULTS} \label{bozomath}
\subsection{Variability}

In order to quantify the intraday variability of the object, we
employ three different statistical methods: $C$ test, $F$ test, the
one-way analysis of variance (ANOVA) (e.g. de Diego 2010; Goyal et
al. 2012; Hu et al. 2014; Dai et al. 2015; Agarwal \& Gupta 2015).
Romero et al. (1999) introduced the variability parameter, $C$, as
the average value between $C_1$ and $C_2$:
\begin{equation}
C_1=\frac{\sigma(BL-StarA)}{\sigma(StarA-StarB)},
C_2=\frac{\sigma(BL-StarB)}{\sigma(StarA-StarB)},
\end{equation}
where (BL-StarA), (BL-StarB) and (StarA-StarB) are the differential
instrumental magnitudes of the blazar and comparison star A, the
blazar and comparison star B, and comparison star A and B. $\sigma$
is the standard deviation of the differential instrumental
magnitudes. The adopted variability criterion requires $C\geq2.576$,
which corresponds to a 99 per cent confidence level. Despite the
very common use of the $C$-statistics, de Diego (2010) has pointed
out that it has severe problems. $F$ test is thought to be a proper
statistics to quantify variability (e.g. de Diego 2010; Joshi et al.
2011; Goyal et al. 2012; Hu et al. 2014; Agarwal \& Gupta 2015). $F$
value is calculated as
\begin{equation}
F_1=\frac{Var(BL-StarA)}{Var(StarA-StarB)},
F_2=\frac{Var(BL-StarB)}{Var(StarA-StarB)},
\end{equation}
where Var(BL-StarA), Var(BL-StarB) and Var(StarA-StarB) are the
variances of differential instrumental magnitudes. The $F$ value
from the average of $F_1$ and $F_2$ is compared with the critical
$F$-value, $F^\alpha_{\nu_{bl},\nu_\ast}$, where $\nu_{bl}$ and
$\nu_{\ast}$ are the number of degrees of freedom for the blazar and
comparison star respectively ($\nu=N-1$), and $\alpha$ is the
significance level set as 0.01 (2.6$\sigma$). If the average $F$
value is larger than the critical value, the blazar is variable at a
confidence level of 99 per cent. de Diego (2010) presented that the
ANOVA is powerful and robust estimator for microvariations. It does
not rely on error measurement but derives the expected variance from
subsamples of the data. The results of de Diego (2010) showed that
according to the choice of one-minute exposures, a group of around
five such exposures, lasting less than 10 minute accounting for CCD
read-out, is a reasonable methodological choice for the ANOVA with
30 minute lags observational strategy. In order to obtain good S/N
ratio, our exposure time were chose from 150 to 400 s. Considering
the time interval of 10 minutes within each group (because 10
minutes is a safe time interval to bin data sharing similar flux
characteristics), we bin the data in a group of three observations.
This method is used only for light curves with more than 9
observations on a given night, but if the measurements in the last
group are less than 3, then it is merged with the previous group.
The critical value of ANOVA can be obtained by
$F^\alpha_{\nu_1,\nu_2}$ in the $F$-statistics, where $\nu_1=k-1$
($k$ is the number of groups), $\nu_2=N-k$ ($N$ is the number of
measurements) and $\alpha$ is the significance level (Hu et al.
2014). Our analysis results on intraday variability are shown in
Table 5. The blazar is considered as variability only when the light
curve satisfies the above three criteria. The blazar is considered
as probably variable if one of the above three criteria is
satisfied. The blazar is considered as non-variable if none of the
criteria are met. From Table 5, it is seen that there is intraday
variability found on 6 nights (2015 April 15, 2014 April 23, 2014
April 24, 2012 April 23, 2012 April 24, 2010 April 03). The
corresponding light curves are seen in Fig. 1. In order to further
quantify the reliability of variability, the $S_x$ is given in Fig.
1. The value of $S_x$ can be calculated as (e.g. Hu et al. 2014)
\begin{equation}
S_x=m_i-\overline{m},~~~x=V,R,I
\end{equation}
where $m_i$ and $\overline{m}$ are same with Equation (1). On 2015
April 15, the $I$ and $R$ bands show intraday variability, and the
$V$ band is probably variable. For the night, the largest magnitude
change of $I$ band is $\bigtriangleup I=0^{\rm m}.25$ in 246 minutes
from MJD=57127.743 to MJD=57127.914. For the $R$ band on this night,
Mrk 501 brightens by $\bigtriangleup R=0^{\rm m}.20$ in 39 minutes
from the beginning of MJD=57127.848 to MJD=57127.875, and then
quickly fades by $\bigtriangleup R=0^{\rm m}.21$ in 13 minutes from
MJD=57127.875 to MJD=57127.884. The largest magnitude change of $R$
band is $\bigtriangleup R=0^{\rm m}.32$ in 52 minutes from
MJD=57127.884 to MJD=57127.920. On 2014 April 23, the magnitude
change of $R$ band is $\bigtriangleup R=0^{\rm m}.15$ in 79 minutes
from MJD=56770.726 to MJD=56770.781. On 2014 April 24, the largest
magnitude change of $I$ band is $\bigtriangleup I=0^{\rm m}.26$ in
141 minutes from MJD=56771.753 to MJD=56771.851. On 2012 April 23,
the largest magnitude change of $I$ band is $\bigtriangleup I=0^{\rm
m}.25$ in 256 minutes from MJD=56040.736 to MJD=56040.914, the
magnitude change of $R$ band is $\bigtriangleup R=0^{\rm m}.11$ in
14 minutes from MJD=56040.763 to MJD=56040.773 and $V$ band
$\bigtriangleup V=0^{\rm m}.13$ in 14 minutes from MJD=56040.730 to
MJD=56040.740. On 2012 April 24, the magnitude change of $I$ band is
$\bigtriangleup I=0^{\rm m}.22$ in 58 minutes from MJD=56041.710 to
MJD=56041.750, $R$ band $\bigtriangleup R=0^{\rm m}.14$ in 43
minutes from MJD=56041.717 to MJD=56041.747 and $V$ band
$\bigtriangleup V=0^{\rm m}.18$ in 14 minutes from MJD=56041.882 to
MJD=56041.892. On 2010 April 03, the largest magnitude change of I
band is $\bigtriangleup I=0^{\rm m}.11$ in 91 minutes from
MJD=55289.787 to MJD=55289.850 and $V$ band $\bigtriangleup V=0^{\rm
m}.14$ in 194 minutes from MJD=55289.763 to MJD=55289.898. The
variability amplitude (Amp) for each night is given in Table 5 and
Fig. 2. The distributions of variability amplitudes in $I$ and $V$
bands are from a few percent to $35\%$ ($R$ band extends to $40\%$),
and most from $10\%$ to $15\%$. The relationships between
variability amplitudes and the source average brightness (see Table
5) are shown in Fig. 3. The correlation between variability
amplitudes and average brightness is found in the three bands (the
coefficients of linear regression analysis $r_{\rm I}=0.33$, $r_{\rm
R}=0.36$, $r_{\rm V}=0.45$), i.e. larger variability amplitude with
a fainter brightness.

The duty cycle (DC) of Mrk 501 is calculated as (Romero et al. 1999;
Stalin et al. 2009; Hu et ai. 2014)
\begin{equation}
{\rm DC}=100\frac{\sum^n_{i=1}N_i(1/\triangle
T_i)}{\sum^n_{i=1}(1/\triangle T_i)}{\rm per~cent},
\end{equation}
where $\triangle T_i=\triangle T_{i,obs}(1+z)^{-1}$, $z$ is the
redshift of the object and $\triangle T_{i,obs}$ is the duration of
the monitoring session of the \emph{ith} night. Note that since for
a given source the monitoring durations on different nights were not
always equal, the computation of DC has been weighted by the actual
monitoring duration $\triangle t_i$ on the \emph{ith} night. $N_i$
will be set to 1 if intraday variability is detected, otherwise
$N_i=0$ (Goyal et al. 2013). When calculating the value of DC, we
only consider nights with more than 9 points and monitoring longer
than one hour. The value of DC is 16.65 per cent for Mrk 501 (54.13
per cent, if PV cases are also included).

Galactic extinction is corrected with the coefficients ($A_{\rm
I}=0^{\rm m}.029$, $A_{\rm R}=0^{\rm m}.041$, $A_{\rm V}=0^{\rm
m}.052$) given in Schlafly \& Finkbeiner (2011) who recalibrated the
results of Schlegel et al. (1998). Following Gaur et al. (2015),
Agarwal \& Gupta (2015) and Aleksic et al. (2015), we subtract the
host galaxy contributions from the observed fluxes in $R$ band by
considering different aperture radii used by different observatories
using the measurements of Nilsson et al. (2007). The host galaxy
contributions in $I$ and $V$ bands are obtained by adopting the
elliptical galaxy colors of $V-R=0.61$, $R-I=0.70$ from Fukugita et
al. (1995). Then we subtract the host galaxy contributions from the
observed fluxes in $I$ and $V$ bands. The magnitudes of $I$, $R$ and
$V$ bands can be converted into fluxes (in mjy) using
$F_I=2550\times 10^{-0.4\ast I}\times10^3$, $F_R=3080\times
10^{-0.4\ast R}\times10^3$ and $F_V=3640\times 10^{-0.4\ast
V}\times10^3$ respectively (Mead et al. 1990). The fluxes of host
galaxy from different aperture radii used by different observatories
are given in Table 2-4. We calculate the error of correcting
magnitude (subtracted the host galaxy contributions) using error
propagation from errors of observed magnitude and host galaxy
magnitude. After correcting Galactic extinction and the host galaxy
contributions, we only use ANOVA to reanalyze the intraday
variability of the object. The blazar is considered as intraday
variability when the light curve satisfies the ANOVA criterion. Then
there are 10 nights detected as intraday variability ($I$ bands of
2015 April 06, 15, May 25, 2014 April 22, and 2012 April 24, $R$
bands of 2015 April 08, 15, 2012 April 23-24 and 2010 April 01, $V$
bands of 2015 May 25, 2014 April 23, 2012 April 23 and 2010 April
03). These nights ($I$ and $V$ of 2015 May 25, $I$ band of 2014
April 22 and $V$ band of 2014 April 23) are though as PV from
results of non-correcting host galaxy contributions because the
light curves satisfy the ANOVA criterion but dissatisfy $C$ and/or
$F$ tests criteria. The three nights ($I$ band of 2015 April 06, $R$
bands of 2015 April 08 and 2010 April 01) are not considered as
intraday variability from non-correcting results. These nights ($I$
bands of 2014 April 24, 2012 April 23 and 2010 April 03, $R$ band of
2012 April 24, $V$ band of 2012 April 24) are detected as intraday
variability from non-correcting results, but not as intraday
variability from correcting results. In order to compare with Fig.
1, a fraction of the light curves (corrected Galactic extinction and
the host galaxy contributions) are given in Fig. 4. The results of
magnitude change are given in Table 6. From Table 6, it is seen that
the results of magnitude change for correcting and non-correcting
Galactic extinction and the host galaxy contributions are different,
and the variability status of partial nights has changed. We take a
variability to be real if the variability is three times greater
than $\sigma$ (Fan et al. 2014). For the non-correcting results, the
largest magnitude change is $\bigtriangleup R=0^{\rm m}.32$ in 52
minutes and the shortest timescale is 13 minutes with
$\bigtriangleup R=0^{\rm m}.21$. For the correcting results, we
obtain that the largest magnitude change is $\bigtriangleup R=0^{\rm
m}.31$ in 72 minutes. The correcting variability amplitude (Amp) for
each night is also given in Fig. 2. The Fig. 2 shows that the
average variability amplitude of correcting magnitude ($<{\rm
Amp}>_I=23.82\%, <{\rm Amp}>_R=22.25\%, <{\rm Amp}>_V=22.05\%$) is
larger than that of non-correcting magnitude. There are still
correlations between variability amplitude (Amp) and average
brightness from correcting results ($r_I=0.67, r_R=0.59, r_V=0.29$;
see Fig. 3). The correcting value of DC is 14.87 per cent excluding
these nights ($I$ and $V$ of 2015 May 25, $I$ band of 2014 April 22
and $V$ band of 2014 April 23), since when calculating the DC of
non-correcting magnitude, these nights are not considered in spite
of meeting ANOVA criterion.

Long-term light curves and color indices variations of our
observations are shown in Fig. 5. From Fig. 5, it is seen that for
correcting and non-correcting results, the source shows variability
and color indices variability on short timescale and long timescale
in the three bands. The overall magnitude changes and color indices
variability are $\bigtriangleup I=0^{\rm m}.75$, $\bigtriangleup
R=0^{\rm m}.62$, $\bigtriangleup V=0^{\rm m}.64$, $\bigtriangleup
(V-R)=0^{\rm m}.4$, $\bigtriangleup (V-I)=0^{\rm m}.47$ and
$\bigtriangleup (R-I)=0^{\rm m}.43$ for non-correcting results, and
$\bigtriangleup I'=1^{\rm m}.53$, $\bigtriangleup R'=1^{\rm m}.33$,
$\bigtriangleup V'=1^{\rm m}.21$ $\bigtriangleup (V'-R')=0^{\rm
m}.7$, $\bigtriangleup (V'-I')=0^{\rm m}.76$ and $\bigtriangleup
(R'-I')=0^{\rm m}.77$ for correcting results. We also use all three
variability tests to quantify the variability of color indices
($V-R$, $V-I$ and $R-I$). On 2010 April 03, the color indices $V-I$
is detected as variability for non-correcting results, but the rest
of nights are not detected as variability for non-correcting
results. After correcting the host galaxy contributions, there are
three nights detected as variability ($V-I$ on 2015 May 25, $V-R$ on
2012 April 23 and $R-I$ on 2010 April 02). The corresponding color
indices variations with time are seen in Fig. 6. On 2012 April 23,
the source shows IDV in $VR$ bands. The result supports that flux
variations are associated with the color variations on intraday
timescale.

\subsection{Cross-correlation analysis and variability timescale}

We perform the inter-band correlation analysis and searched for the
possible inter-band time delay. First, the $V$ band magnitude versus
$I$ band magnitude is given in Fig. 7. The Fig. 7 and linear
regression analysis show that there are strong correlations between
the $V$ band magnitude and $I$ band magnitude. The interpolated
cross-correlation function (ICCF; Gaskell \& Peterson 1987) and the
discrete correlation function (DCF; Edelson \& Krolik 1988) are
standard techniques in time series analysis for finding time lags
between light curves at different wavelengths. Apart from the ICCF
and DCF, there is another method of estimating the cross-correlation
function (CCF) in the case of non-uniformly sampled light curves,
the z-transformed discrete correlation function (ZDCF; Alexander
1997). It has been shown in practice that the calculation of the
ZDCF is more robust than that of the ICCF and the DCF when applied
to sparsely and unequally sampled light curves (e.g., Edelson et al.
1996; Giveon et al. 1999; Roy et al. 2000). The ZDCF is used in our
analysis. The ZDCF code of Alexander (1997) can automatically set
how many bins are given and used to calculate the inter-band
correlation on intraday and long time scales. In order to keep the
validity of the statistical data, we set minimal ten number of
points per bin. For magnitude of non-corrected host galaxy
contributions, the part of the results are displayed in Fig. 8. For
each correlation, a Gaussian fitting is made to find the central
ZDCF points. The time where the Gaussian profile peaks denotes the
lag between the correlated light curves (Wu et al. 2012). From Fig.
8 and results of Gaussian fitting, and considering time resolutions
and spans, we can find that there are not significant time lags
between the $V$ band magnitude and $I$ band magnitude. For some
nights and long time scales, we have not enough data points to find
time lags because most of nights have less than 20 data points and
there is large observed time interval for long time scale. The
correcting results from ZDCF analysis are consistent with
non-correcting results.

The zero-crossing time of the autocorrelation function (ACF) is a
well-defined quantity and is used as a characteristic variability
timescale (e.g. Alexander 1997; Giveon et al. 1999; Netzer et al.
1996; Liu et al. 2008). If there is an underlying signal in the
light curve, with a typical timescale, then the width of the ACF
peak near zero time lag will be proportional to this timescale
(Giveon et al. 1999; Liu et al. 2008). Following Giveon et al.
(1999) and Liu et al. (2008), we choose the zero-crossing time of
the ACF as the variability timescale. In addition, the first-order
structure function (SF; Trevese et al. 1994) can also be used to
analyze a characteristic variability timescale. There is a simple
relation between the ACF and the SF (see equation (8) in Giveon et
al. 1999). In this paper, only an ACF analysis is performed to
search for the presence of any characteristic variability timescale.
We estimate the ACF also using the ZDCF code from Alexander (1997),
and only analyze the nights detected as intraday variability. After
correcting the host galaxy contributions, the results used to
estimate characteristic variability timescale are given in Fig. 9.
We then used a least-squares procedure to fit a fifth-order
polynomial to the ZDCF, with the constraint that ACF($\tau$=0)=1
(Giveon et al. 1999). From Fig. 9 and fitting results, intraday
variation timescales of $\sim23$ to $\sim171$ minutes were detected.
The most common definition of the variability timescale
$\tau=F/(|\triangle F/\triangle t|)$ has the advantage of weighting
fluctuations by their amplitudes, where $F$ is the flux density at
the minimum, and $\triangle F$ is the variability amplitude in the
timescale $\triangle t$ (e.g. Wagner \& Witzel 1995; Liu et al.
2008). The variation timescale is defined as variations of
$\triangle F/F\geq30\%$ between the subsequent minimum and maximum
within the timescale $\triangle t$, and $\triangle F> 3\sigma$ (Liu
et al. 2008; Fan et al. 2014). In order to further confirm
reliability of the variation timescales, we check the corresponding
light curves, and find that 2012 April 23 $R'$ (from 56040.783 to
56040.822), 2012 April 24 $R'$ (from 56041.757 to 56041.707) and
2015 April 15 $I'$ (from 57127.743 to 57127.915) satisfy the
criteria of the variation timescales (see Fig. 4). Therefore, from
our results, three characteristic timescales 0.0737 0.0752, and
0.119 days are detected. Among them, the 0.0737 day $\sim$106 min
from 2012 April 23 is the minimal characteristic variability
timescale.

\subsection{Correlation between Magnitude and Color Index}

In order to explore the optical spectral properties, we analyze the
correlation between magnitude and color index for intraday,
short-term, and long-term timescales. The relation between $V-R$
index and $V$ magnitude is frequently studied, so we concentrate on
$V-R$ index and $V$ magnitude. When exploring the correlation for
intraday timescale, we only analyze the nights with the number of
$V-R$ index $N\geq9$. For the color index, first, we correct the
Galactic extinction. After correcting the Galactic extinction, the
results of correlations between $V-R$ index and $V$ magnitude are
given in Table 7. Then the results of correcting the host galaxy
contributions are also given in Table 7. As an example, Fig. 10
shows the correlations between $V-R$ index and $V$ magnitude for
non-correcting the host galaxy contributions on intraday timescale.
From Fig. 10 and Table 7, it is seen that most of nights have strong
correlations between $V-R$ index and $V$ magnitude on intraday
timescale for non-correcting and correcting the host galaxy
contributions. There are three nights with moderate correlations for
correcting or non-correcting the host galaxy contributions (2010
April 01, 2012 April 24 and 2013 April 03). So a bluer-when-brighter
(BWB) chromatic trend is dominant for Mrk 501 on intraday timescale.
And the BWB trend exists in active and stable state because in our
observations, some nights show IDV, while others do not. There is
still a strong BWB trend on short-term timescale (2010 April 01-03,
2012 Apr 22-24 and 2013 Apr 01-04 from Table 7). In a
intermediate-term timescale (weeks to months), the correlations
weaken (see Table 7). For long-term timescale, the correlations for
correcting the host galaxy contributions are stronger than that for
non-correcting the host galaxy contributions (2014 Apr 19-May 08 and
2015 Apr 06-May 25 from Table 7). The correlations between $V-R$
index and $V$ magnitude in long-term timescales are shown in Fig.
11. The analysis of spearman rank obtains that there is a weak
correlations between $V-R$ index and $V$ magnitude for
non-correcting the host galaxy contributions ($r=0.1, P=0.06$). For
all data subtracted the host galaxy contributions, there is
significant negative correlation between $V-R$ index and $V$
magnitude ($r=-0.235, P=1\times 10^{-5}$), i.e. redder-when-brighter
(RWB). However, if we inspect the Fig. 11 in detail, we can find two
separate regions (a) and (b). The dividing line is $V\sim14^{\rm
m}.55$. For region (a), there is significant correlation with
$V-R\sim0.48(\pm0.08)V$ ($r=0.38, P=1\times10^{-6}$) and significant
correlation with $V-R\sim0.29(\pm0.06)V$ ($r=0.35,
P=9\times10^{-6}$) for region (b). The correlation coefficient and
slope of region (a) are larger than that of region (b), i.e.
compared with low flux state, the BWB behavior is more pronounced at
high flux state.

\section{DISCUSSION AND CONCLUSIONS} \label{bozomath}

Most blazars show microvariability/IDV in a few hours/minutes, but
the unpredictability of these changes in brightness and the
difficulty of obtaining confirmation by others have been an
incredible cause (de Diego 2010). Generally, $C$ test is used for
quantifying the IDV. de Diego (2010) presented that $C$ test is not
a reliable methodology, and that $F$ test and ANOVA are powerful and
robust estimators for IDV. In order to increase the confidence on
the validity of IDV, the blazar is considered as IDV only when the
light curve satisfies all the three criteria. For ANOVA, The results
of de Diego (2010) showed that a binning of five observations is a
reasonable methodological choice for the ANOVA with 30 minute lags
observational strategy. But considering our exposure time, we bin
the data in a group of three observations. Note that the smallest
bin sizes improve the time resolutions, but the detections are
biased toward large amplitude (de Diego 2010). The CCD aperture
photometry obtains the optical magnitude by comparing the object
brightness to the brightness of several comparison stars in the
field of the object which causes that changes in sky transparency
can be eliminated and accuracies $<1\%$ attained even under varying
conditions (Nilsson et al. 2007). However, the presence of a strong
host galaxy component can affect measured aperture photometry. Mrk
501 has strong host galaxy components. If the host galaxy
contribution are not subtracted, the broadband spectra are highly
distorted and relative optical flux variations underestimated. FWHM
changes do not introduce very strong false variability in objects
with a prominent host galaxy component (0.01-0.03 mag; Cellone et
al. 2000; Nilsson et al. 2007). For our observations, The aperture
radius of 2$\times$FWHM or 1.5$\times$FWHM was selected, and the
scale of aperture radius is same on the same night. The value of
FWHM is associated with seeing. One night, the value of FWHM almost
is constant, but the small seeing change can still cause changes of
FWHM and aperture radius. These can bring in false variability in
objects with a prominent host galaxy component. For a longer-term
timescale, different nights or periods are likely to have larger
seeing change which may cause larger false variability in objects
with a prominent host galaxy component. Therefore, for the objects
with a prominent host galaxy component, the correcting host galaxy
components are needed. We subtract the host galaxy contributions
from the observed fluxes in $R$ band by considering different
aperture radii used by different observatories using the
measurements of Nilsson et al. (2007). The host galaxy contributions
in $I$ and $V$ bands are not obtained from direct measurements but
from conversion of colors. This may introduce uncertainty.

Long-term quasi-simultaneous multi-band observations of blazar Mrk
501 were performed from 2010 to 2015 in $V$, $R$ and $I$ bands. Our
monitoring program for Mrk 501 was divided into five different
periods. In total, we observed 29 nights and 1186 CCD frames with
error less than 0.05 mags. For our results of non-correcting host
galaxy contributions, the magnitude distributions in $V$, $R$ and
$I$ bands are $14^{\rm m}.14<V<13^{\rm m}.50$, $13^{\rm
m}.71<R<13^{\rm m}.08$ and $13^{\rm m}.07<I<12^{\rm m}.32$, and mean
values $<V>=13^{\rm m}.85\pm0^{\rm m}.12$, $<R>=13^{\rm
m}.38\pm0^{\rm m}.12$ and $<I>=12^{\rm m}.74\pm0^{\rm m}.14$. The
magnitude distributions in $VRI$ bands in 1995-1996 from Xie et al.
(1999) are $14^{\rm m}.16-13^{\rm m}.79$, $13^{\rm m}.72-13^{\rm
m}.20$ and $13^{\rm m}.17-12^{\rm m}.62$ respectively. The magnitude
distribution in $R$ band in 1999 from Xie et al. (2001) is from
$13^{\rm m}.04$ to $12^{\rm m}.96$. The magnitude distribution in
$R$ band in 2000 from Dai et al. (2001) is from $13^{\rm m}.77$ to
$13^{\rm m}.09$. The magnitude distributions in $VR$ bands in 2000
from Gupta et al. (2008) are $13^{\rm m}.77-13^{\rm m}.59$ and
$13^{\rm m}.45-13^{\rm m}.17$ respectively. The magnitude
distribution in $R$ band in 2007-2012 from Wierzcholska et al.
(2015) is from $13^{\rm m}.57$ to $13^{\rm m}.37$. The magnitude
distribution in $V$ band in 2011-2012 from Shukla et al. (2015) is
from $14^{\rm m}.05$ to $13^{\rm m}.87$. From above references and
NASA/IPAC Extragalactic Database: NED, we obtain that the maximum
and minimum magnitude for this source till data are $14^{\rm
m}.16-13^{\rm m}.39$ (Xie et al. 1999; Hutter \& Mufson 1981),
$13^{\rm m}.77-12^{\rm m}.64$ (Dai et al. 2001; Hutter \& Mufson
1981) and $13^{\rm m}.17-12^{\rm m}.21$ (Xie et al. 1999; Sitko et
al. 1983) for $V$, $R$ and $I$ bands respectively. Through
comparisons, it is seen that our distributive ranges of magnitude in
$VRI$ bands are basically consistent with others.

\subsection{Variability}

Optical photometric observations of blazars are important tools for
constructing their light curves which can yield valuable information
about the mechanisms operating in these sources, with important
implications for blazars modeling (e.g. Fan et al. 2009).
Variability timescale is one of the observational characteristics of
blazars. Detailed knowledge of the statistical behaviors on
different timescales is therefore very useful to understand the
basic physical mechanisms during the flaring and quiescent phases
(Dai et al. 2015; Ciprini et al. 2003).

From our analysis, for results of non-correcting host galaxy
contributions, there is intraday variability found on 6 nights. The
distributions of variability amplitudes in $I$ and $V$ bands are
from a few percent to $35\%$ ($R$ band extends to $40\%$), and most
from $10\%$ to $15\%$. The value of DC is 16.65 per cent for Mrk 501
(54.13 per cent, if PV cases are also included). Goyal et al. (2013)
found that for TeV blazars the DC is $45\%$ and if IDV with
variability amplitudes ${\rm Amp}>3\%$ is considered, the DC is
$32\%$. Our average variability amplitude is much larger than that
of TeV blazars from table 1 of Goyal et al. (2013). So compared with
other TeV blazars, the Mrk 501 may have large variability amplitude
and low DC. Making use of results of Nilsson et al. (2007), we
subtract the host galaxy contributions from the observed fluxes. The
correcting results show that the results of magnitude change for
correcting and non-correcting the host galaxy contributions are
different, and the variability status of partial nights has changed.
These are mainly because the seeing change causes changes of FWHM
and aperture radius. After correcting the host galaxy contributions,
the variability status and magnitude variabilities will change. The
average variability amplitude of correcting magnitude is larger than
that of non-correcting magnitude, which supports that the relative
optical flux variations are underestimated for results of
non-correcting host galaxy contribution. If the optical flux
variations include object and host galaxy components, and host
galaxy is more likely to be as stable component, after subtracting
host galaxy component, the variability amplitude will increase. The
correcting value of DC is slightly smaller than the non-correcting
value. The results of Xie et al. (1996) showed that intraday
variability amplitude of Mrk 501 is smaller than 0.22 mag in the $B$
and $V$ bands. From results of Xie (1999, 2001), Dai et al. (2001),
Gupta et al. (2008), Wierzcholska et al. (2015) and Shukla et al.
(2015), for Mrk 501, the intraday variability amplitude with more
than 0.3 mag is rare. Long-term light curves of our observations
show that the overall magnitude changes are $\bigtriangleup I=0^{\rm
m}.75$, $\bigtriangleup R=0^{\rm m}.62$ and $\bigtriangleup V=0^{\rm
m}.64$ for non-correcting results, and $\bigtriangleup I=1^{\rm
m}.53$, $\bigtriangleup R=1^{\rm m}.33$ and $\bigtriangleup V=1^{\rm
m}.21$ for correcting results. Xie et al. (1999) presented that Mrk
501 show the largest variation $\triangle V=0^{\rm m}$.8 during 345
days. From Xie (1999, 2001), Dai et al. (2001), Gupta et al. (2008),
Wierzcholska et al. (2015) and Shukla et al. (2015), the variability
amplitude in long-term timescale from $0^{\rm m}.2$ to $0^{\rm m}.7$
are observed. From results of ACF and analysis of light curves, a
minimal variability timescale of $\sim106$ min in $\triangle
R'=0.31$ is detected. Ghosh et al (2000) found that the observed
variability of Mrk 501 by 0.13 mag within 12 minutes. Xie et al.
(1999) presented that the Mrk 501 has a flare of $\Delta R=0.49$ in
105 minutes. Gupta et al. (2008) found that the Mrk 501 has a flare
of $\Delta R\sim0.05$ in 15 minutes. So our minimal variability
timescale is consistent with result from Xie et al. (1999). If the
observed minimum timescale indicates an innermost stable orbital
period, an upper limit can be obtained for the mass of the central
black hole (e.g. Xie et al. 2004; Fan et al. 2014; Dai et al. 2015).
From Fan (2005) and Fan et al. (2014), the innermost stable orbit
depends on the black hole and the accretion disk, and $r\leq
c\delta\triangle t_{\rm min}/(1+z)$, and then the upper limits of
black hole masses are (i) $M_8\leq1.2\times\frac{\delta\triangle
t_{\rm min}(hr)}{1+z}$ for a thin accretion disk surrounding a
Schwarzschild black hole; (ii)
$M_8\leq1.8\times\frac{\delta\triangle t_{\rm min}(hr)}{1+z}$ for a
thick accretion disk surrounding a Schwarzschild black hole; (iii)
$M_8\leq(\frac{7.3}{1+\sqrt{1-a^2}})(\frac{\delta\triangle t_{\rm
min}(hr)}{1+z})$ for the case of a Kerr black hole, where $M_8$ is
the black hole mass in units of $10^8M_\odot$, and $a$ is an angular
momentum parameter. Then we obtain the upper limits of black hole
mass are $M\leq10^{8.99}M_\odot$ and $M\leq10^{9.17}M_\odot$ for a
thin and thick accretion disks, and $M\leq10^{9.78}M_\odot$ for the
extreme Kerr black hole case ($a=1$) with Doppler factor 4.8 from Wu
et al. (2007). Considering the minimal variability timescale as the
light crossing time of the emitting volume (e.g. Dai et al. 2015),
the size of the emitting region $r\leq8.9\times10^{14}$ cm is also
estimated. Woo \& Urry (2002) and Sbarrato et al. (2012) estimated
virial black hole mass for AGNs. The black hole masses of Mrk 501
from Woo \& Urry (2002) and Sbarrato et al. (2012) are
$10^{9}M_\odot$ and $10^{9.21}M_\odot$ respectively. Therefore, our
upper limits of black hole mass estimated from variability timescale
is consistent with results of Woo \& Urry (2002) and Sbarrato et al.
(2012).

In order to explain the optical IDV, many intrinsic and extrinsic
models have been proposed. The intrinsic models include shock-in-jet
model (Marscher \& Gear 1985; Marscher et al. 2008) and accretion
disc model (Chakrabarti \& Wiita 1993; Mangalam \& Wiita 1993).
Extrinsic mechanisms involve interstellar scintillation and
geometrical effects (Heeschen et al. 1987; Gopal-Krishna \& Wiita
1992). The interstellar scintillation leads to radio variations at
low frequencies and can therefore not be the case of optical
intraday variability (Agarwal \& Gupta 2015). The long-term periodic
and achromatic BL Lacertae variability may be mostly explained by
the geometrical scenarios where viewing angle variation can be due
to the rotation of an inhomogeneous helical jet which causes
variable Doppler boosting of the corresponding radiation (e.g. Gaur
et al. 2015; Larionov et al. 2010, Villata et al. 2002). Among the
several accepted theoretical models explaining variability, for Mrk
501 in high state, the origin of IDV and STV can be attributed to
the shock-in-jet model. The accretion disc models are not likely to
explain the IDV because the accretion disk radiation is always
overwhelmed by the Doppler boosting flux from the relativistic jet,
and Mrk 501 has a radiatively inefficient accretion discs (Urry \&
Padovani 1995; Ghisellini et al. 2011; Sbarrato et al. 2012; Xiong
et al. 2014). However, the blazars in low state, the origin of IDV
can be explain by instability in the accretion disk, because in the
low state, jet emission is less dominant over the emission from the
disk. A weak emission from base of jet is an alternative way to
explain the IDV of Mrk 501 in the low state (Gupta et al. 2008).
Moreover, for high SMBH masses, the ligh crossing time is very large
which can cause the variability of shortest timescale (Agarwal \&
Gupta 2015). On basis of the synchrotron self-Compton (SSC)
mechanism, Shukla et al. (2015) presented that the observed optical
and X-ray variability of Mrk 501 may be explained by the injection
of fresh electrons in emission zones and cooling of the electrons.

For Mrk 501, one of the most fascinating features is the observed
possible periodicity of $\sim$ 23 days in the TeV and X-ray band
(e.g. Hayashida et al. 1998; kranich et al. 1999). Rieger \&
Mannheim (2000) presented that this periodicity possibly be related
to the orbital motion of the relativistic jet in a binary black hole
systems where the nonthermal radiation is emitted by a relativistic
jet which emerges from the less massive black hole, the periodicity
thus being due mainly to geometrical origin (i.e. Doppler-shifted
modulation). The origin of the periodicity of $\sim$ 23 days may be
explained by the instabilities of (optically thin) accretion disks
(Fan et al. 2008). From our observations, we can not obtain possible
periodic variability timescales for STV and LTV because there are
not enough observed data points.

\subsection{Relation of Color and Magnitude} \label{bozomath}

For Mrk 501, the host galaxy contributions and Galactic extinction
can lead to a unreal color variations. From our results, the BWB
trend is dominant for Mrk 501 on intermediate, short and intraday
timescales for both non-correcting and correcting host galaxy
contributions. Compared with short timescale, the long timescale has
a weaker BWB trend, and there is a weak BWB trend in the long
timescale for non-correcting results. The long timescale are likely
to have larger seeing change which may cause larger false
variability in Mrk 501. So the weak BWB trend may be due to the
contamination of host galaxy. In the case of non-correcting host
galaxy contributions, Wierzcholska et al. (2015) and Ikejiri et al.
(2011) have found significant overall BWB trend in long timescale
for Mrk 501. We should note that the largest variability amplitudes
of Mrk 501 from them are less than $0^{\rm m}.2$. Zheng et al.
(2008) found that flux variations are associated with the color
variations in long term timescale. Then the color variations may be
related with active state of the object. For the results of them, it
may stand for relatively quiet state. From our results, after
correcting the host galaxy contributions, there is a overall RWB
trend in the long timescale. It is possible that the RWB trend is
due to superposition of different components (jet components,
accretion disk components, Gravitational microlensing effect;
Bonning et al. 2012; Fan et al. 2008; Gaur et al. 2015; Agarwal \&
Gupta 2015). The uncertainties of correcting host galaxy components
may also cause the RWB trend. However, we can find two separate
regions in the long-term timescale. The dividing line is
$V\sim14^{\rm m}.55$. The two separate regions show the BWB trend
but have different slope of BWB. Therefore, for long-term timescale,
Mrk 501 in different state can have different BWB trend, and
compared with low flux state, the BWB behavior is more pronounced at
high flux state.

The BWB trends is very well known for BL Lacs. It is important to
recall that the underlying host galaxy, effect of the accretion disc
component and the Gravitational microlensing can also lead to
apparent, but unreal, color variations (Gaur et al. 2015; Hawkins
2002). Since Mrk 501 has a radiatively inefficient accretion discs;
our results exclude the host galaxy contributions; Gravitational
microlensing is important on weeks to months timescales, the source
is a good candidate for studying color variations for intraday
timescale. The BWB trend indicates ``harder when brighter'' behavior
which is consistent with the shock-in-jet model (e.g. Dai et al.
2015). According to the shock-in-jet model, as the shock propagating
down the jet strikes a region of high electron population,
radiations at different visible colors is produced at different
distances behind the shocks. High-energy photons from synchrotron
mechanism typically emerge sooner and closer to the shock front than
the lower frequency radiation thus causing color variations (Agarwal
\& Gupta 2015). From our results, the BWB trends on intermediate,
short and intraday timescales support the shock-in-jet model. For
long-term timescale, host galaxy effect, Doppler factor variation
(Villata et al. 2004; Hu et al. 2014) and superposition of different
components (Bonning et al. 2012; Gaur et al. 2015) may cause the
weak BWB (non-corrected host galaxy contributions) or RWB trends
(corrected host galaxy contributions). Moreover, for long-term
timescale, the different states with different BWB trends support
the result from Anderhub et al. (2009) and Acciari et al. (2011)
that compared with low-activity state, the ``harder when brighter''
behavior are more pronounced at high-activity level. Finally, we
should note that for color variations, simultaneous multi-band
observations are needed.

The results of ZDCF analysis indicate that there are not significant
time lags between the $V$ band magnitude and $I$ band magnitude.
Generally, on short timescales, it is difficult to detect the time
lags between two optical bands due to the closeness of the various
optical bands (Gaur et al. 2015). We also found the smaller
variability amplitude with a larger brightness which can be
explained as the source flux increases, the irregularities in the
turbulent jet decrease and the jet flow becomes more uniform leading
to a decrease in amplitude of variability (Marscher 2014; Gaur et
al. 2015).

\section{SUMMARY}

We have monitored the BL Lac object Mrk 501 in five periods from
2010 to 2015. The quasi-simultaneous multi-band observations in
$VRI$ bands provide 1186 data points with error less than 0.05 mags.
Our main results are the following:

(i) For results of non-correcting host galaxy contributions, there
is intraday variability found on 6 nights. The distributions of
variability amplitudes in $I$ and $V$ bands are from a few percent
to $35\%$ ($R$ band extends to $40\%$), and most from $10\%$ to
$15\%$. The value of DC is 16.65 per cent for Mrk 501 (54.13 per
cent, if PV cases are also included). Long-term light curves of our
observations show that the overall magnitude changes are
$\bigtriangleup I=0^{\rm m}.75$, $\bigtriangleup R=0^{\rm m}.62$ and
$\bigtriangleup V=0^{\rm m}.64$.

(ii) Compared with non-correcting results, after correcting host
galaxy contributions, the magnitude change are different, and the
IDV status of partial nights has changed. The average variability
amplitude of correcting magnitude are larger than that of
non-correcting magnitude. The correcting value of DC is slightly
smaller than the non-correcting value. The overall magnitude changes
are $\bigtriangleup I=1^{\rm m}.53$, $\bigtriangleup R=1^{\rm m}.33$
and $\bigtriangleup V=1^{\rm m}.21$.

(iii) There are inverse correlations between the variability
amplitude and the average brightness both for non-correcting and
correcting results. No significant time lag between $V$ and $I$
bands is found on one night. A minimal variability timescale of 106
minutes is detected.

(iv) The source shows color indices variability on intraday
timescale, short timescale and long timescale. The flux variations
are associated with the color variations on intraday timescale.

(v) The BWB trend is dominant for Mrk 501 on intermediate, short and
intraday timescales for both non-correcting and correcting host
galaxy contributions which supports the shock-in-jet model. For the
long timescale, Mrk 501 in different state can have different BWB
trend.

\begin{acknowledgements}
We sincerely thank the referee for valuable comments and
suggestions. DRX also thanks Benzhong Dai and Wei Zeng for help.
This work is financially supported by the National Nature Science
Foundation of China (11163007, U1231203, 11063004, 11133006 and
11361140347) and the Strategic Priority Research Program ``The
emergence of Cosmological Structures'' of the Chinese Academy of
Sciences (grant No. XDB09000000). This research has made use of the
NASA/IPAC Extragalactic Database (NED), that is operated by Jet
Propulsion Laboratory, California Institute of Technology, under
contract with the National Aeronautics and Space Administration.
\end{acknowledgements}

\begin{figure}
\begin{center}
\includegraphics[width=20cm,height=20cm]{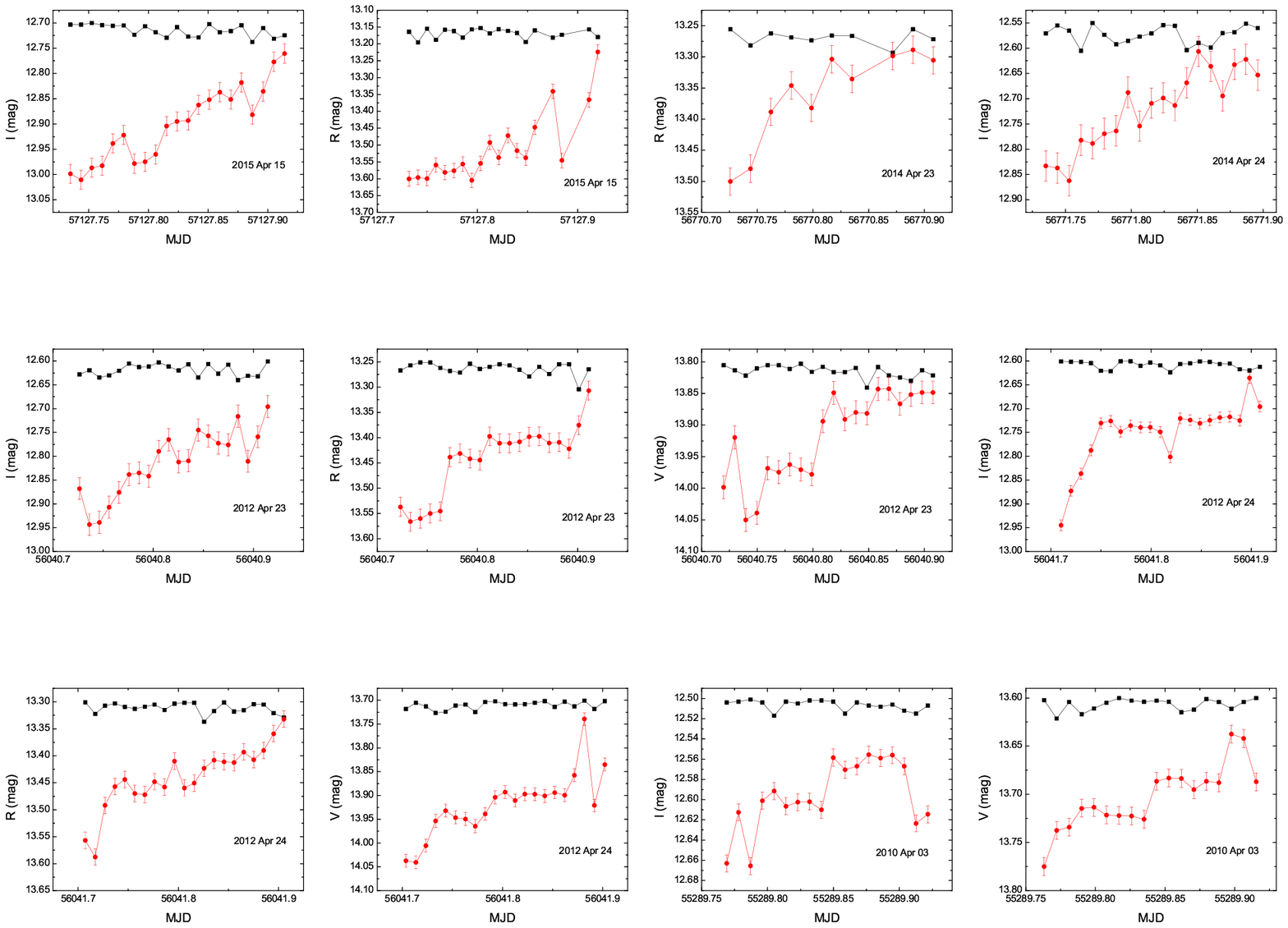}
\caption{Light curves of intraday variability for Mrk 501. The red
circles and lines are the light curves for Mrk 501. The black
squares and lines are the variations of $S_I$, $S_R$ and $S_V$. The
light curves of $S_I$, $S_R$ and $S_V$ are offset to avoid their
eclipsing with light curves of Mrk 501. \label{fig3}}
\end{center}
\end{figure}

\begin{figure}
\begin{center}
\includegraphics[angle=0,width=0.65\textwidth]{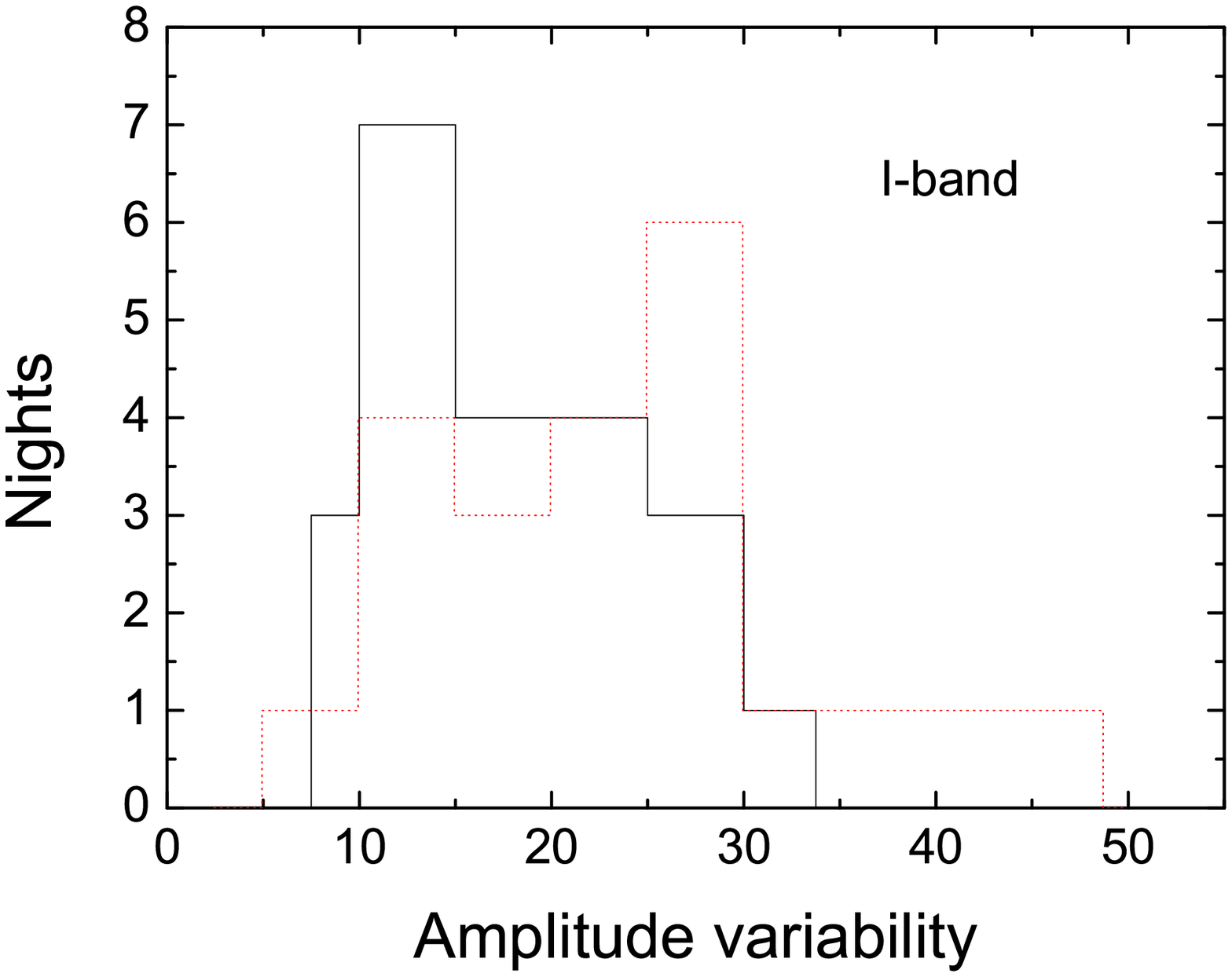}
\includegraphics[angle=0,width=0.65\textwidth]{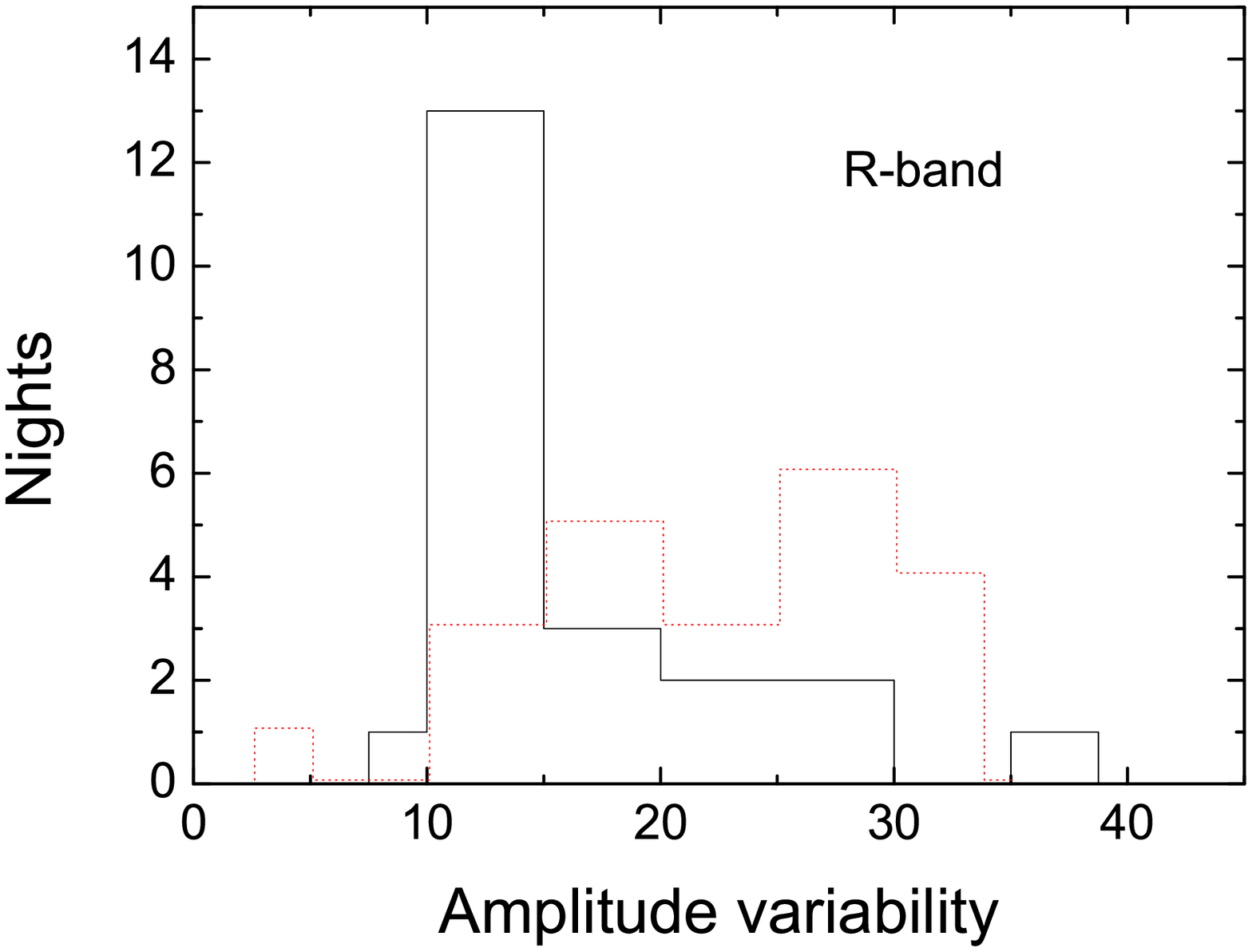}
\includegraphics[angle=0,width=0.65\textwidth]{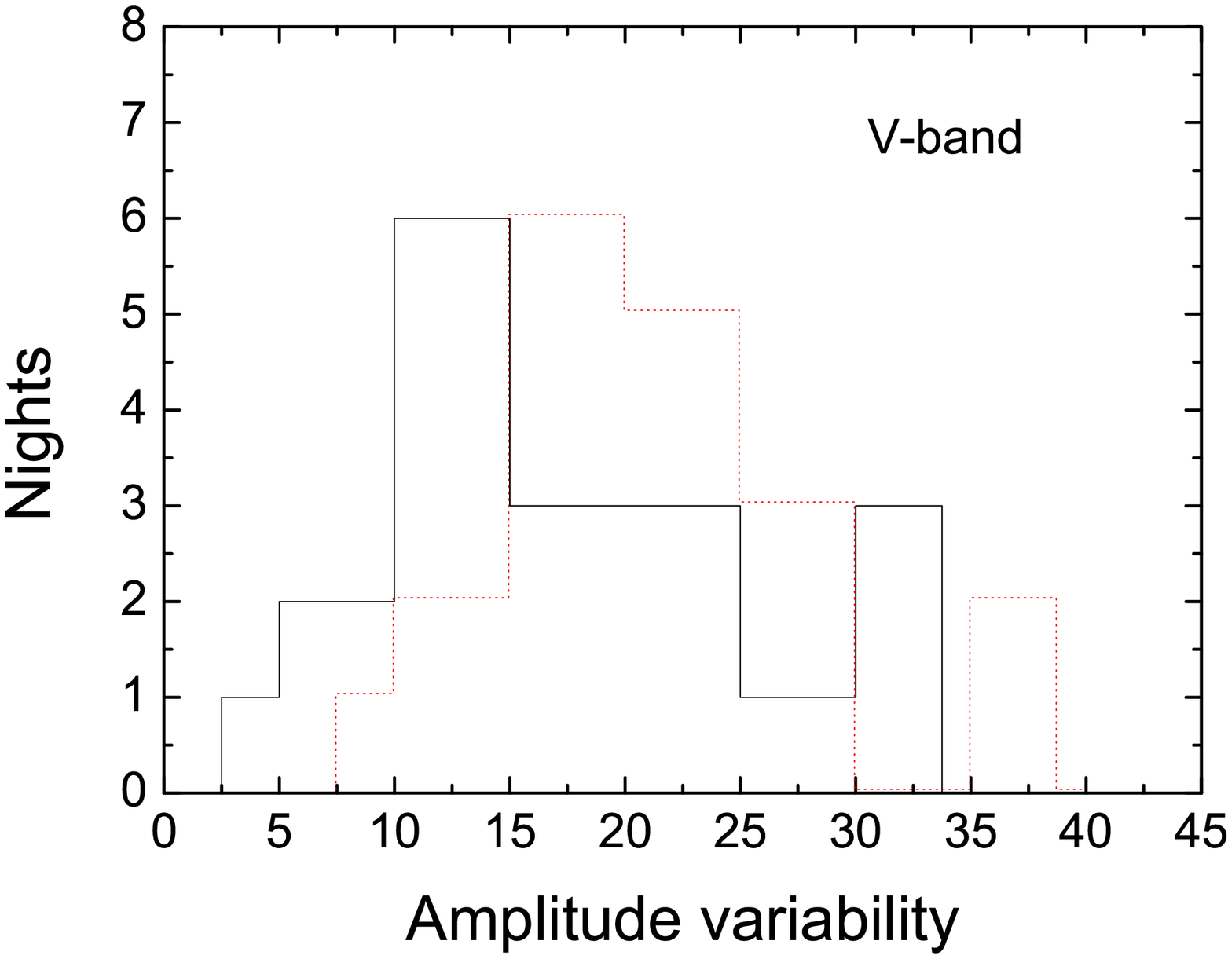}
\caption{The intraday variability amplitude distribution in $I$, $R$
and $V$ bands. The black solid lines stand for non-correcting
results and the red dotted lines for results of correcting Galactic
extinction and the host galaxy contributions.}
\end{center}
\end{figure}

\begin{figure}
\begin{center}
\includegraphics[angle=0,width=0.65\textwidth]{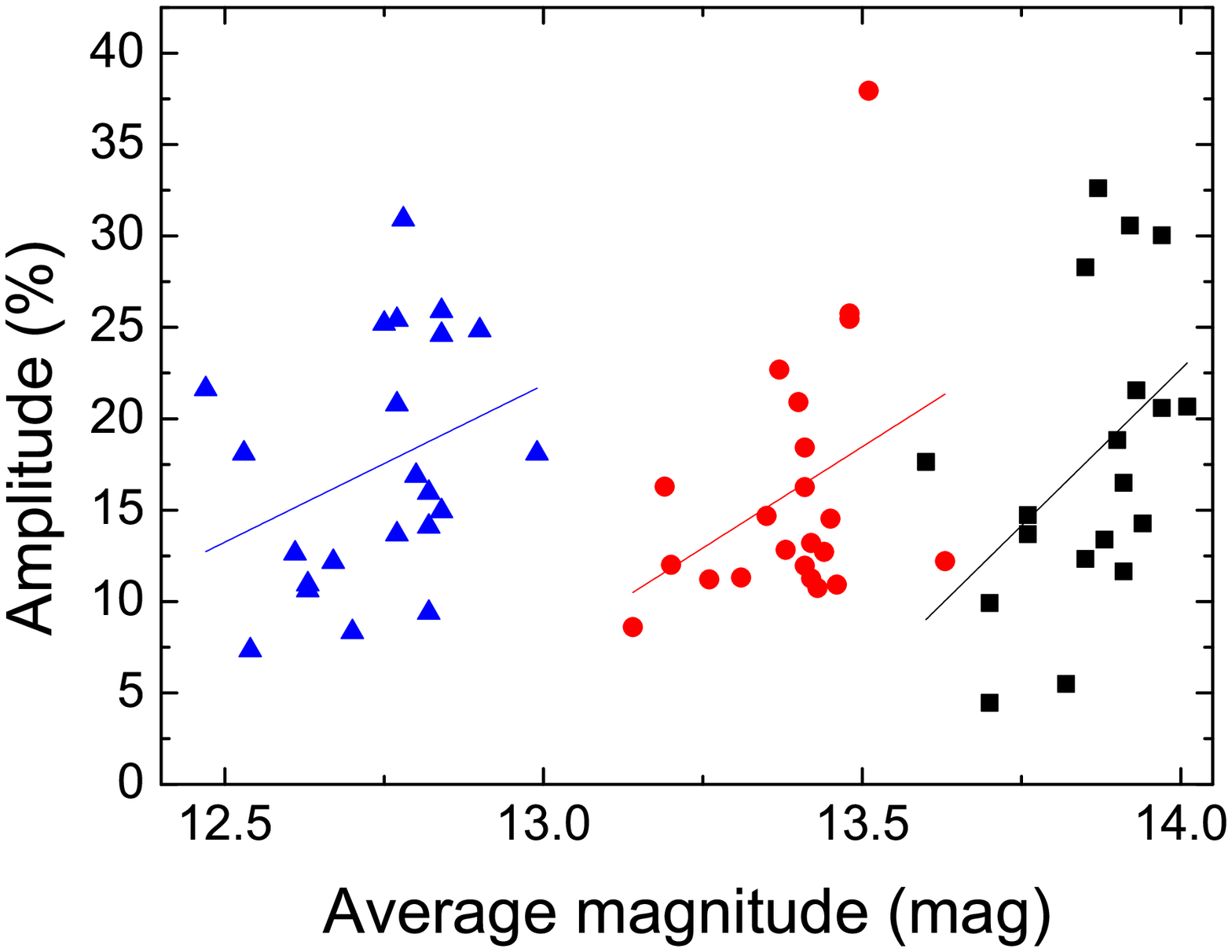}
\includegraphics[angle=0,width=0.65\textwidth]{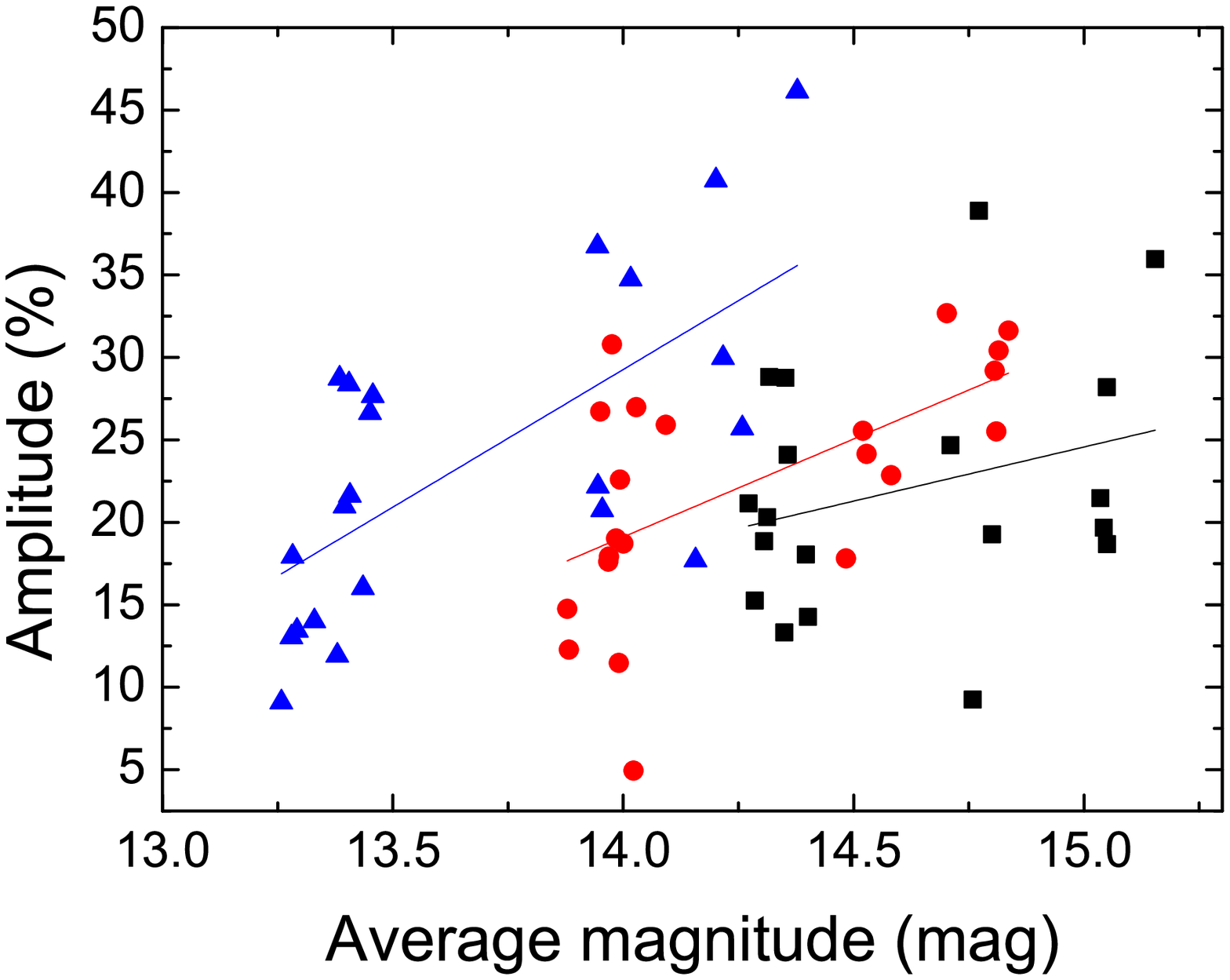}
\caption{The variability amplitude versus the average brightness.
The top panel is the results of non-correcting Galactic extinction
and the host galaxy contributions, and the bottom panel is the
correcting results. The blue triangles, red circles and black
squares stand for $I$, $R$ and $V$ bands respectively. The lines are
the best linear fitting for the three bands.}
\end{center}
\end{figure}

\begin{figure}
\begin{center}
\includegraphics[width=25cm,height=20cm]{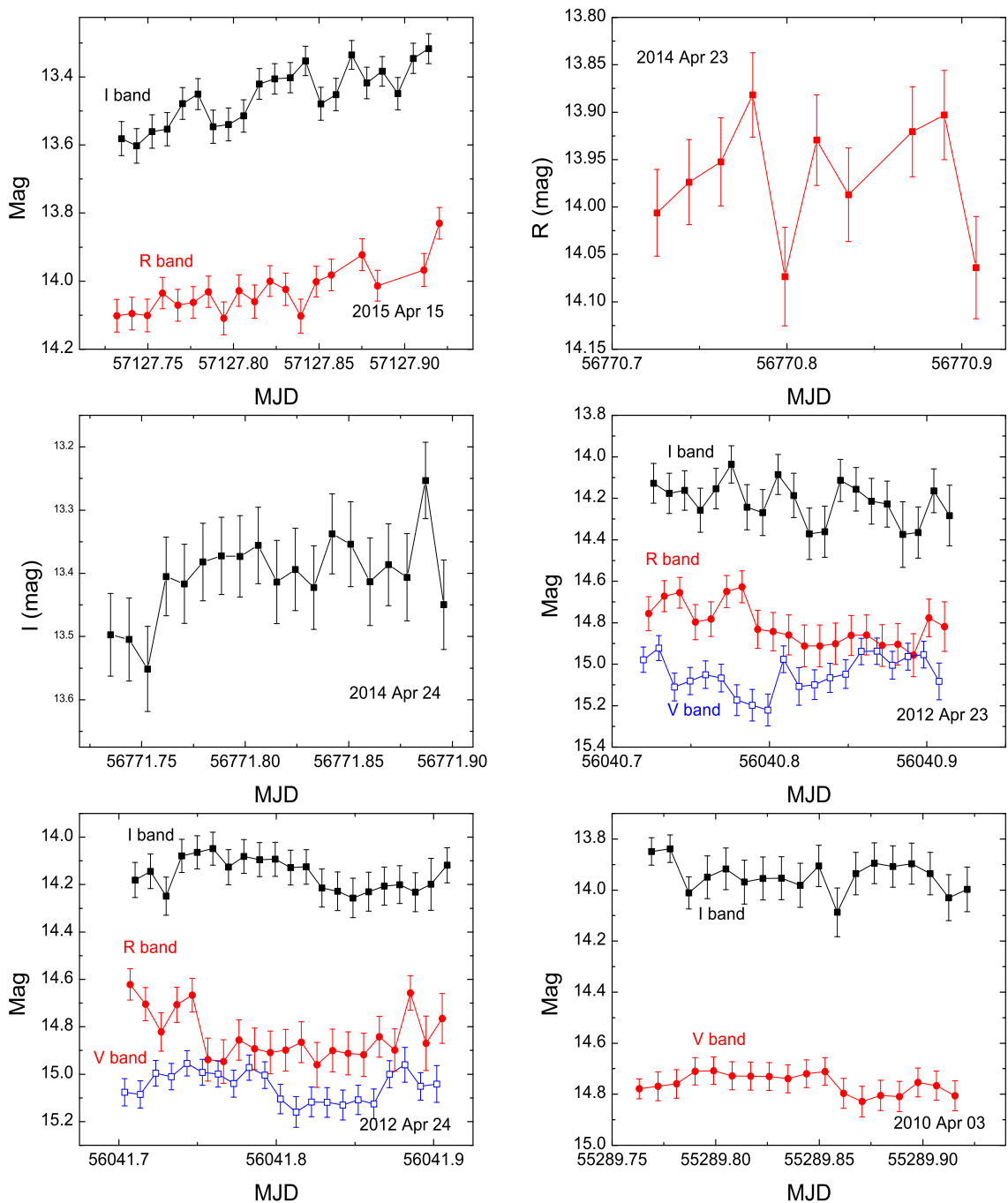}
\caption{Light curves (corrected Galactic extinction and the host
galaxy contributions) for Mrk 501. The black filled squares, red
filled circles and blue empty squares are $I$, $R$ and $V$ bands
respectively. \label{fig5}}
\end{center}
\end{figure}

\begin{figure}
\begin{center}
\includegraphics[width=20cm,height=20cm]{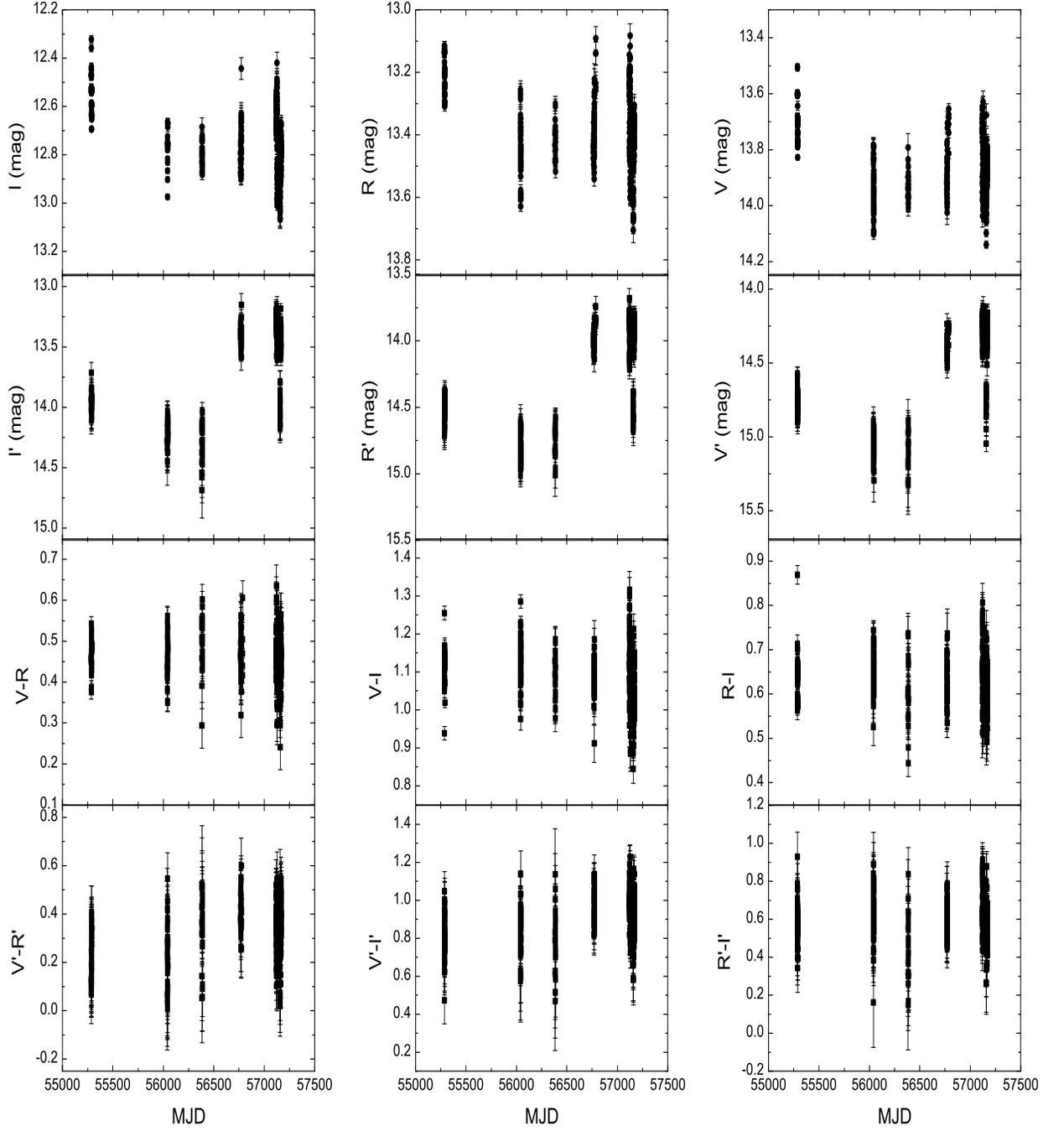}
\caption{Long-term light curves of Mrk 501 in $I$, $R$, $V$ bands
and color indices $V-R$, $V-I$ and $R-I$. The first and third rows
of panels are the results of non-correcting host galaxy
contributions, and the second and forth rows of panels are the
results of correcting host galaxy contributions\label{fig6}}
\end{center}
\end{figure}

\begin{figure}
\begin{center}
\includegraphics[width=16cm,height=16cm]{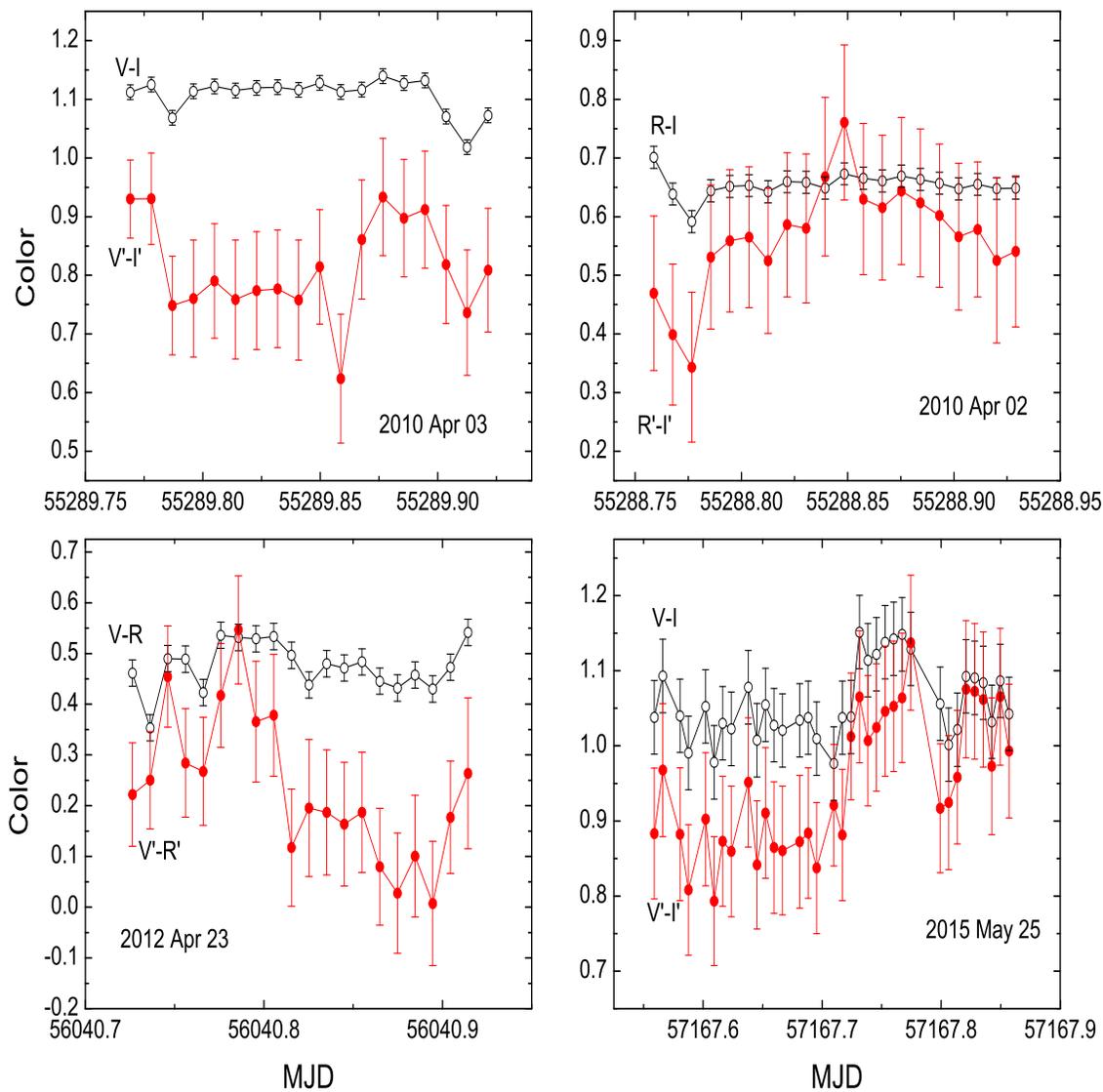}
\caption{Color indices variations with time for intraday timescale.
The black empty circles stands for color indices of non-correcting
host galaxy contributions, and red filled circles for color indices
of correcting host galaxy contributions. \label{fig6}}
\end{center}
\end{figure}

\begin{figure}
\begin{center}
\includegraphics[width=20cm,height=18cm]{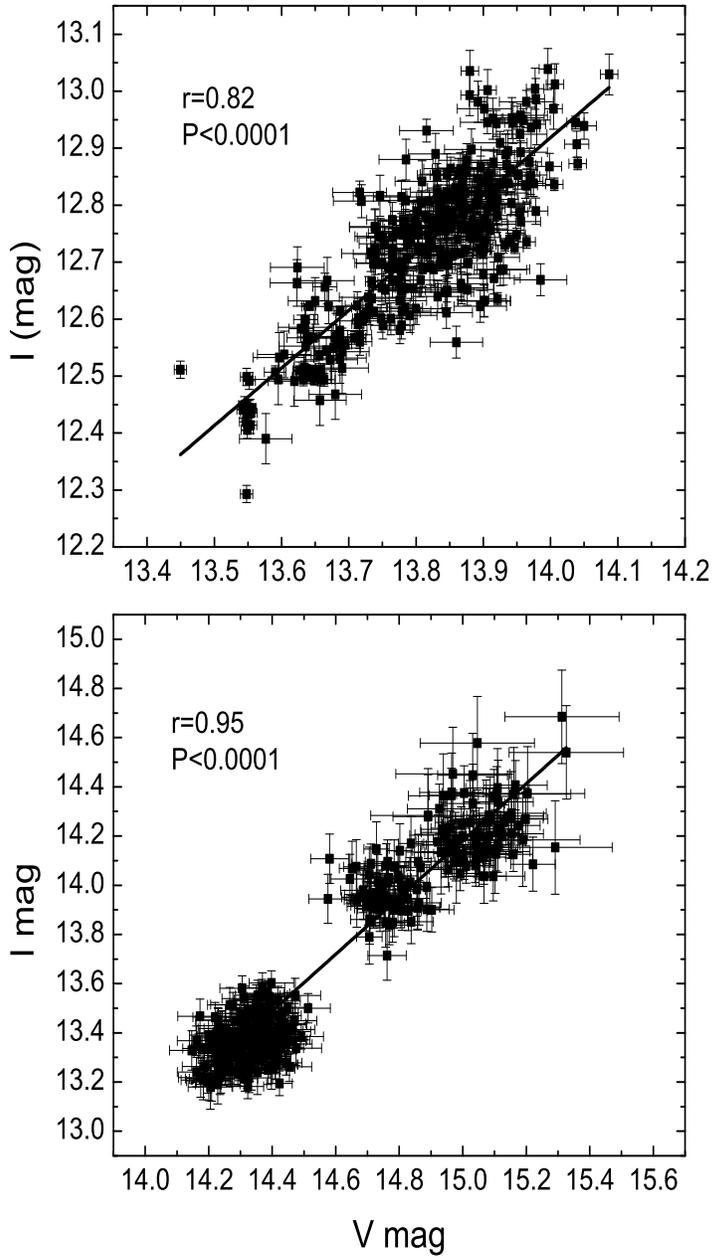}
\caption{The $I$ band magnitude versus $V$ band magnitude for
non-correcting (top panel) and correcting (bottom panel) host galaxy
contributions. The solid lines are the results of linear regression
analysis. $r$ is the correlation coefficient; $P$ is the chance
probability.\label{fig6}}
\end{center}
\end{figure}

\begin{figure}
\begin{center}
\includegraphics[angle=0,width=0.48\textwidth]{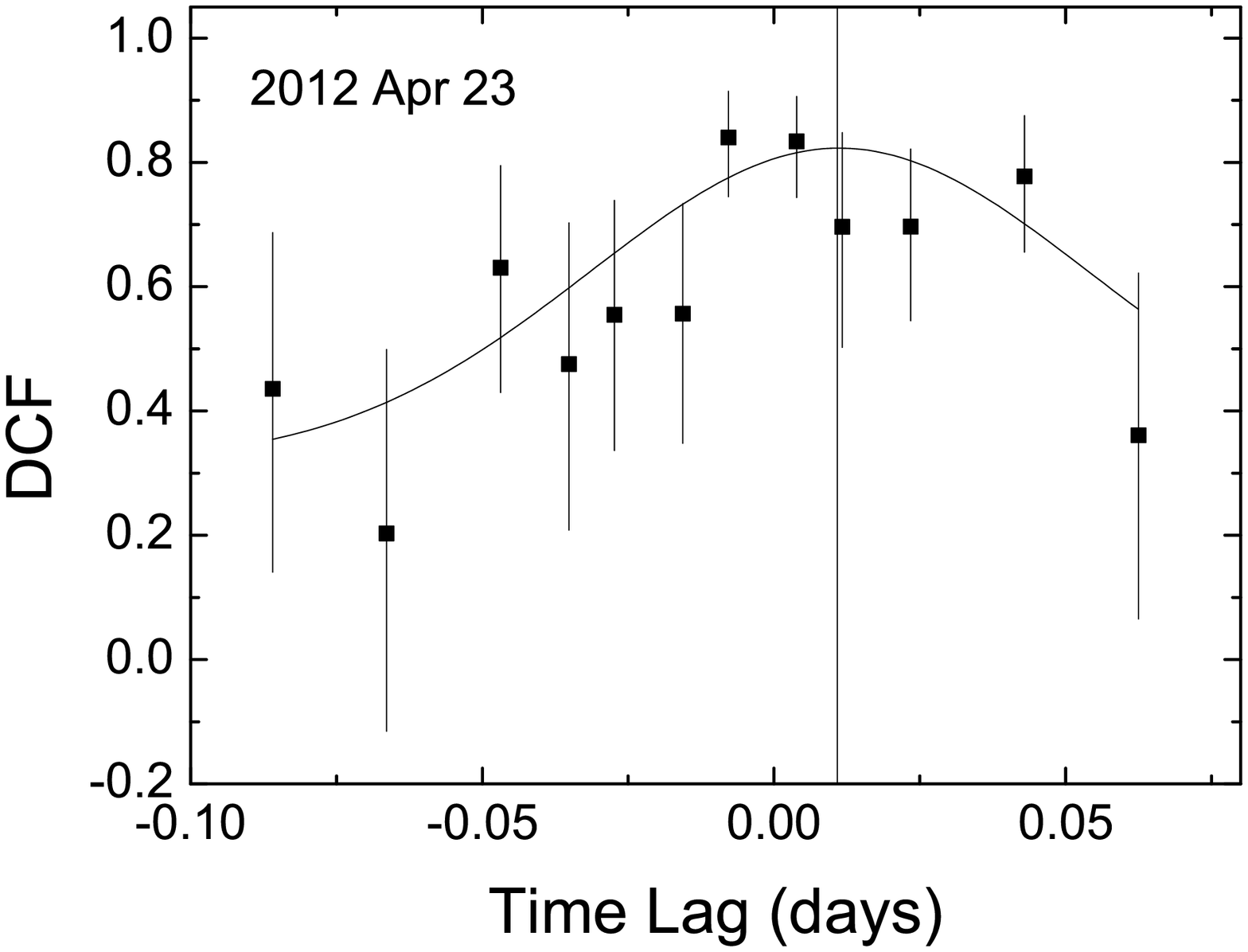}
\includegraphics[angle=0,width=0.48\textwidth]{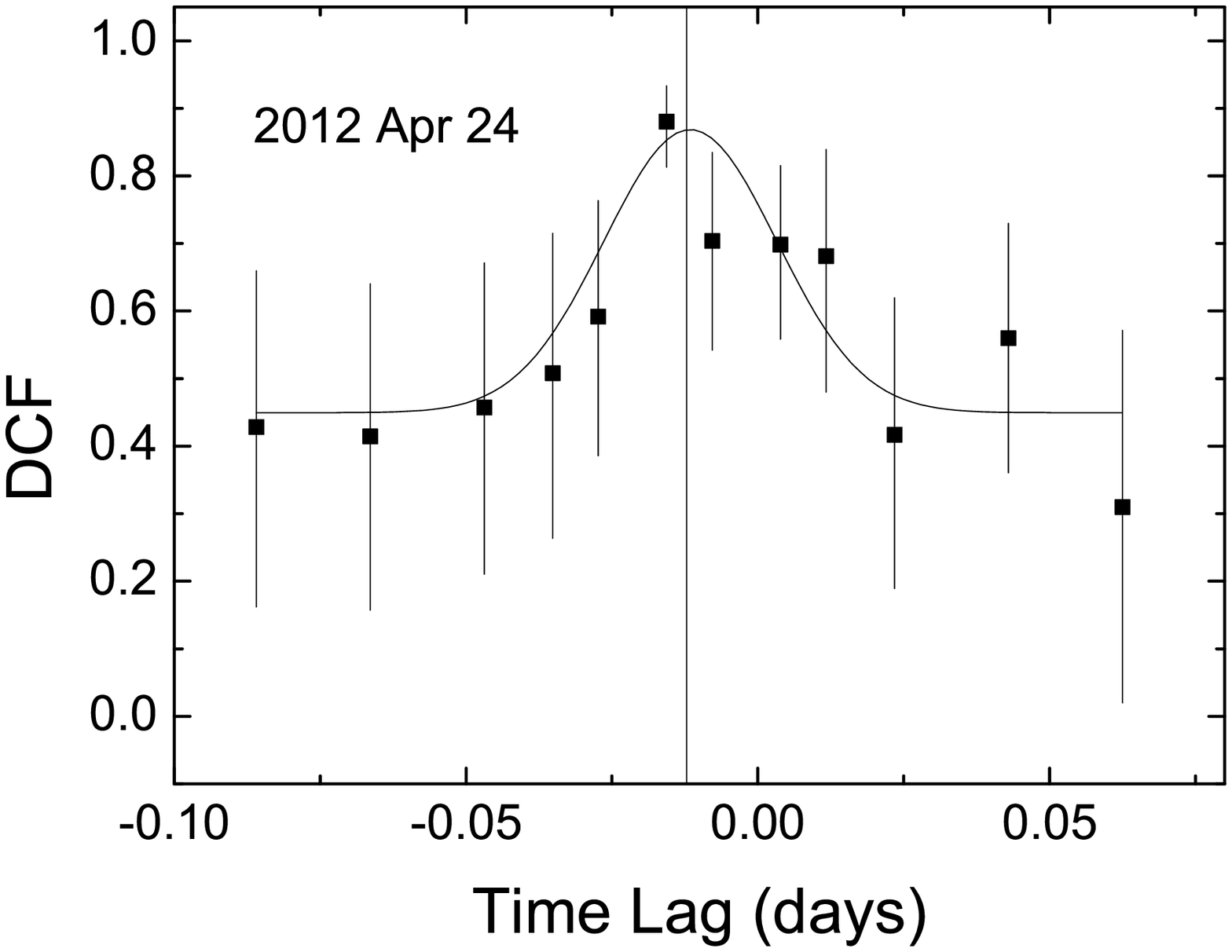}
\includegraphics[angle=0,width=0.48\textwidth]{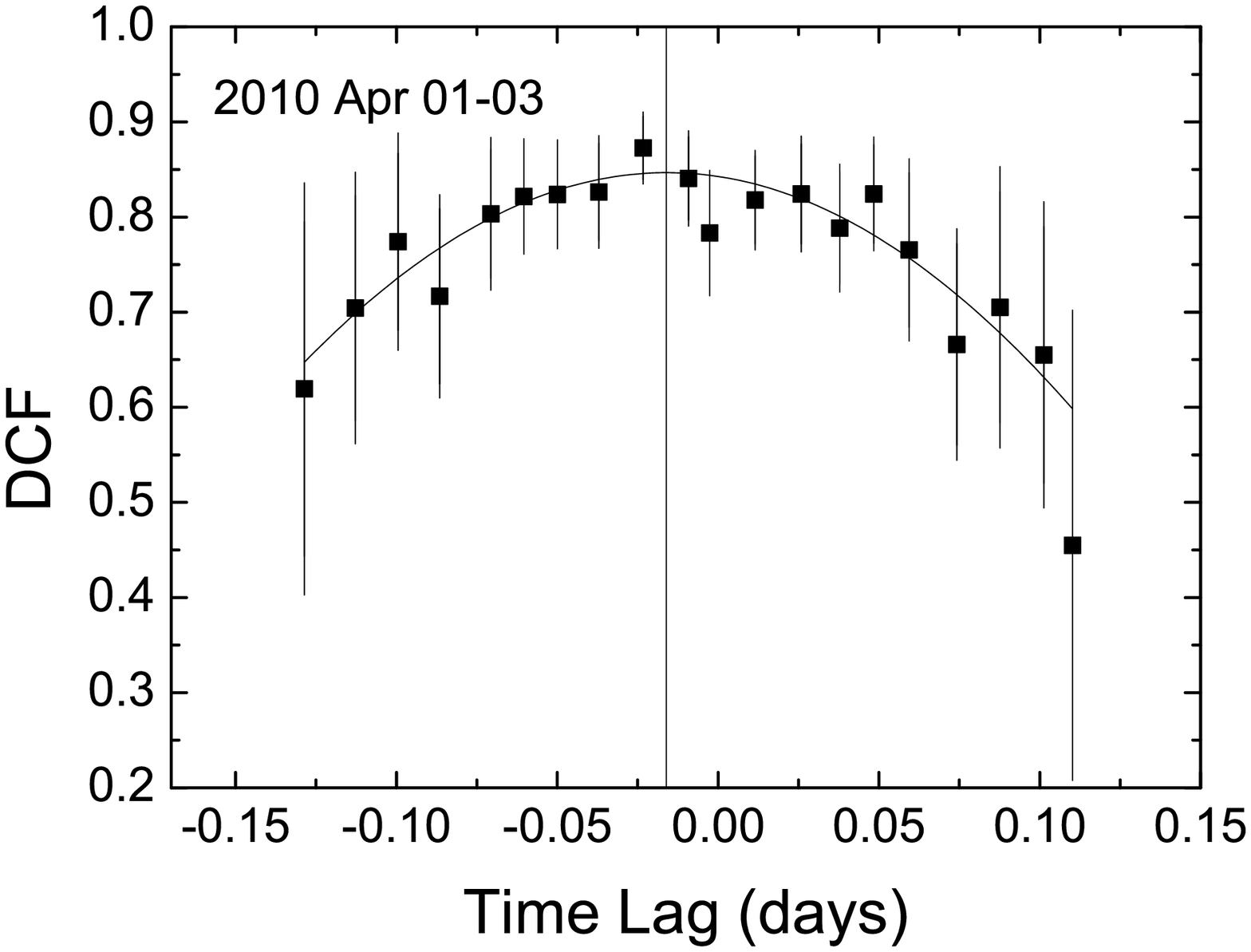}
\includegraphics[angle=0,width=0.48\textwidth]{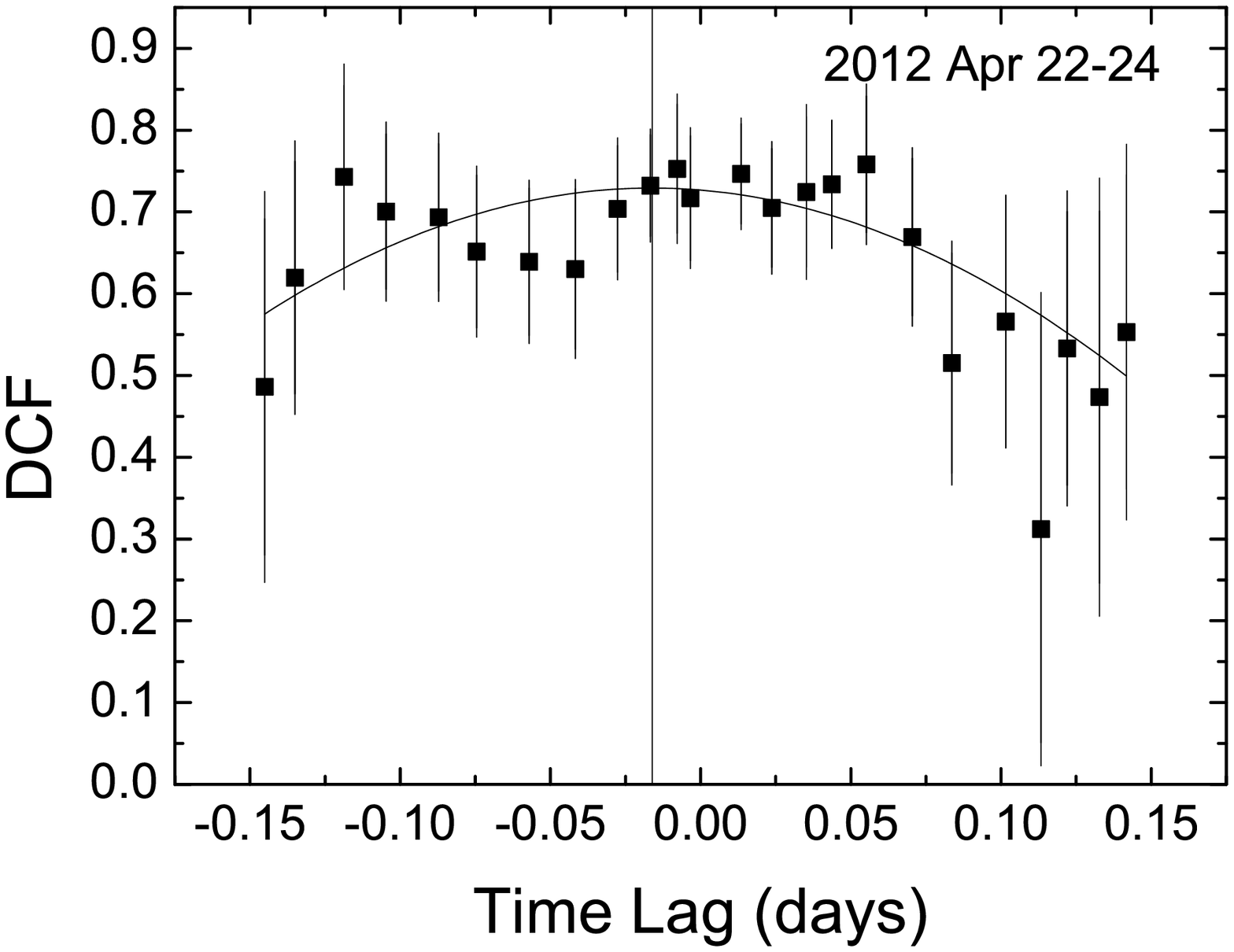}
\includegraphics[angle=0,width=0.48\textwidth]{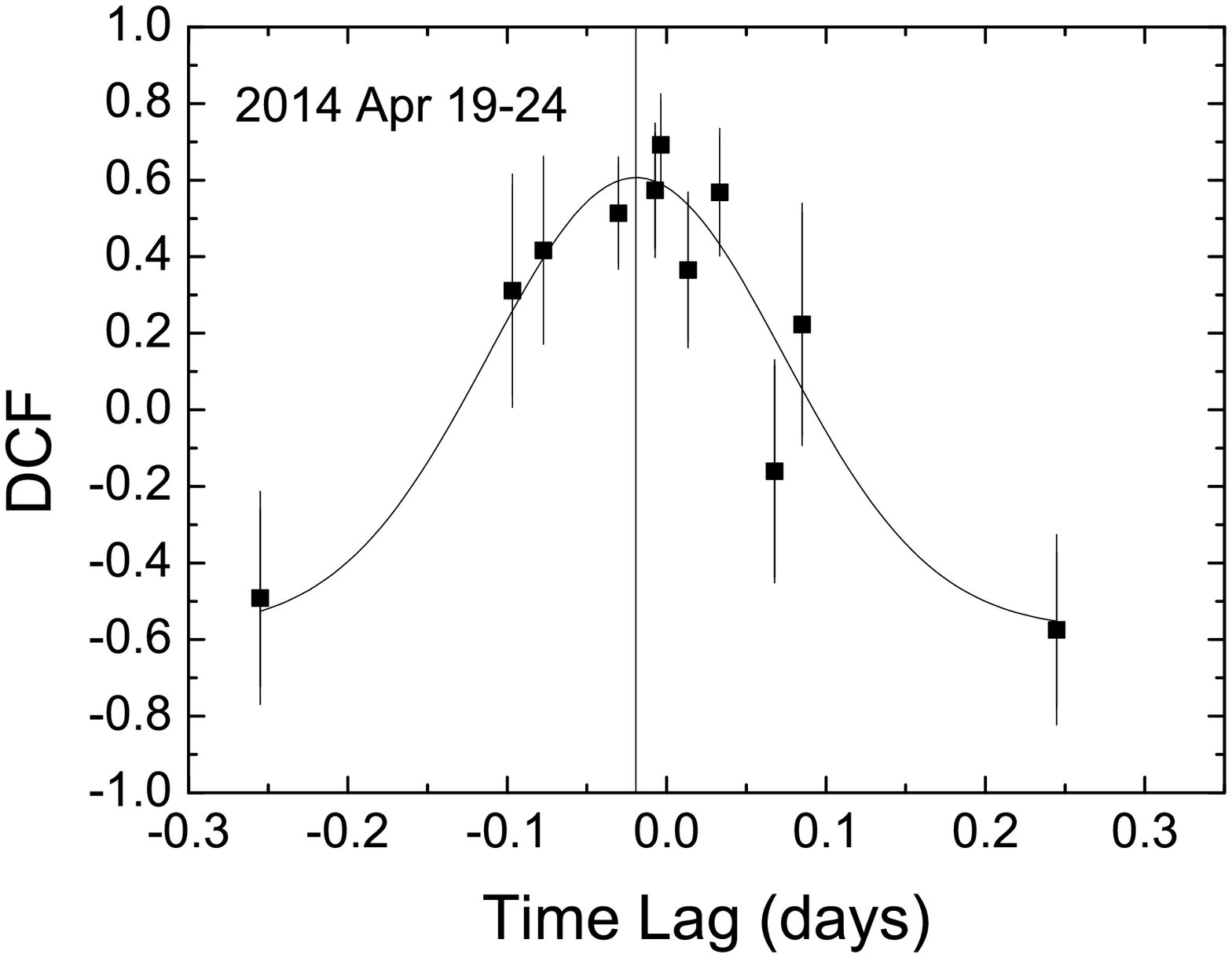}
\includegraphics[angle=0,width=0.48\textwidth]{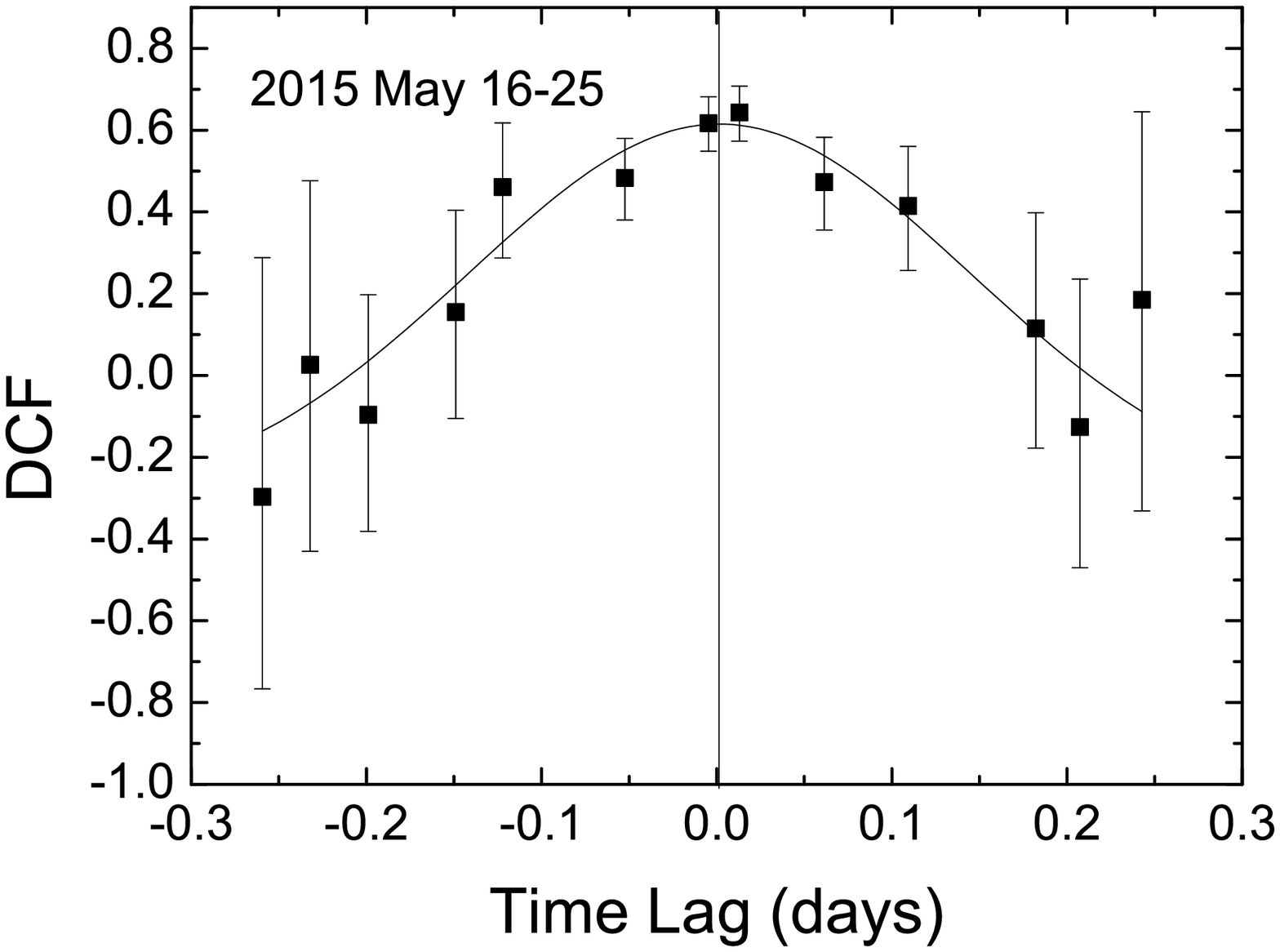}
\caption{The ZDCF plots between $V$ and $I$ bands. The curves show
Gaussian fittings to the points, and their peaks are marked with the
vertical lines. The corresponding time lags are $0.011\pm0.007,
-0.012\pm0.003, -0.016\pm0.005, -0.016\pm0.011, -0.019\pm0.012$ and
$0.0016\pm0.01$ in days for 2012 April 23, 2012 April 24, 2010 April
01-03, 2012 April 22-24, 2014 April 19-24 and 2015 May 16-25.}
\end{center}
\end{figure}

\begin{figure}
\begin{center}
\includegraphics[width=22cm,height=22cm]{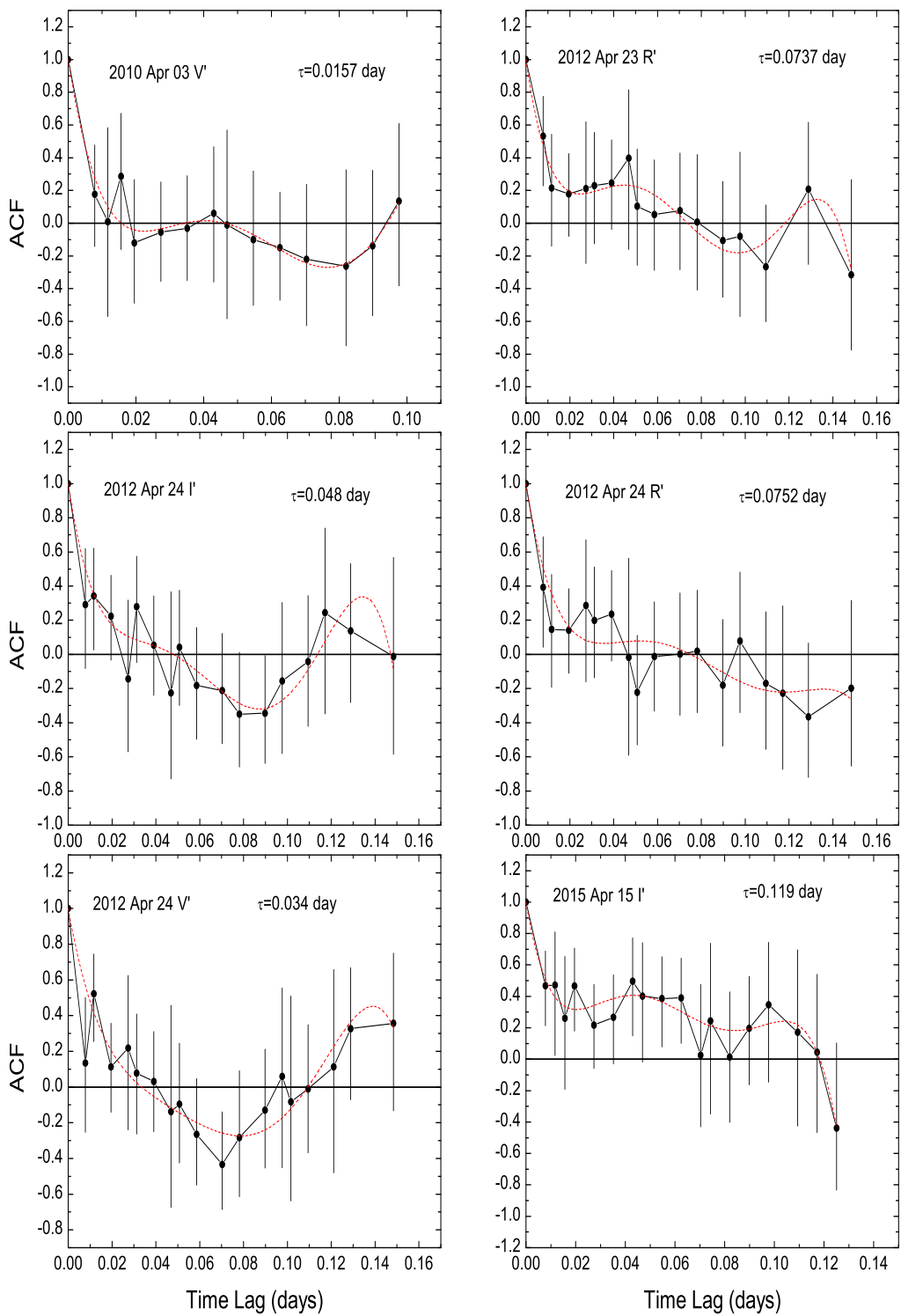}
\caption{The ACF analysis for results of correcting the host galaxy
contributions. The red dash line is a fifth-order polynomial
least-squares fit.\label{fig3}}
\end{center}
\end{figure}

\begin{figure}
\begin{center}
\includegraphics[width=20cm,height=22cm]{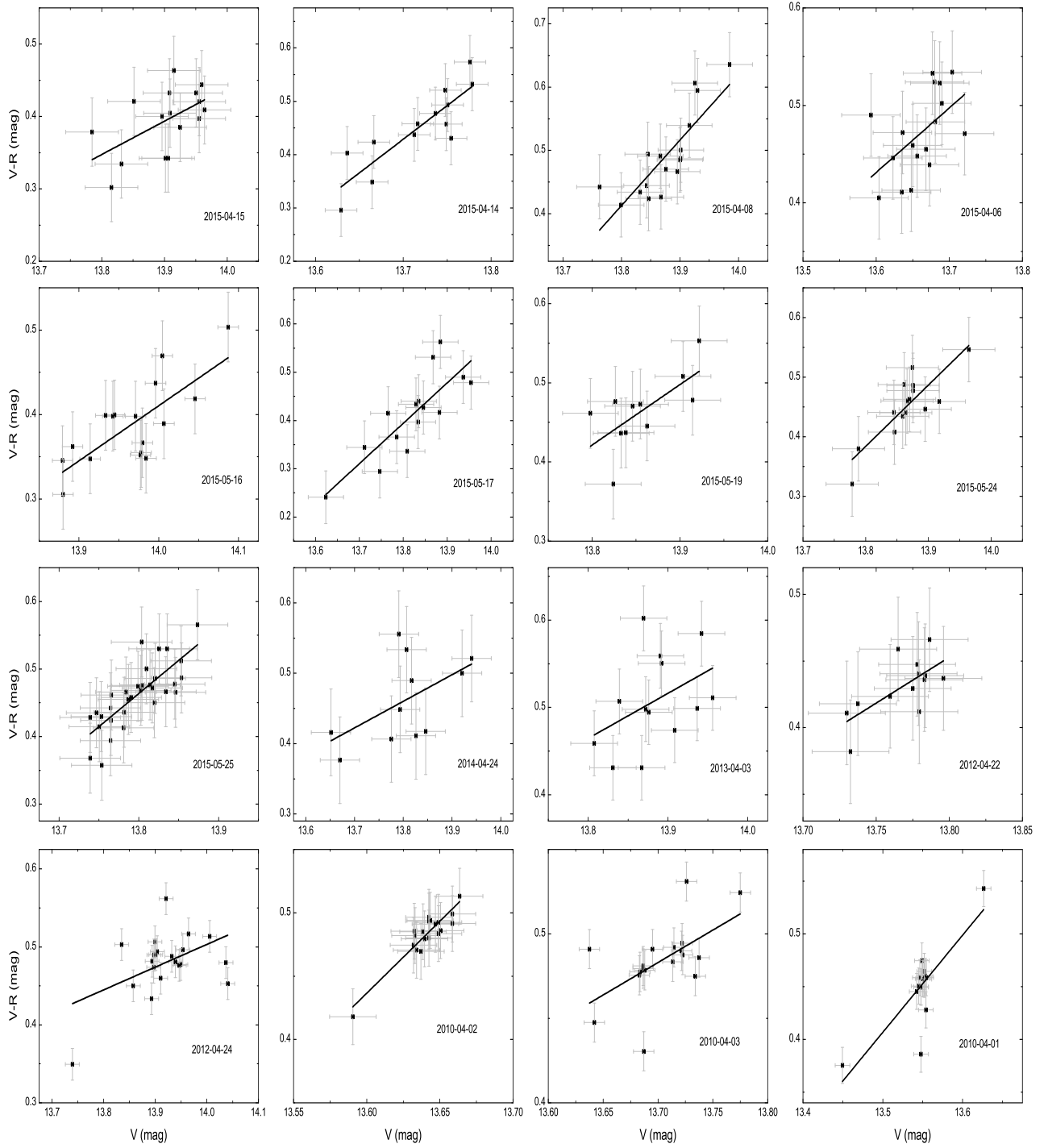}
\caption{The correlations between $V-R$ index and $V$ magnitude for
non-correcting the host galaxy contributions. The solid lines is
results of linear regression analysis. \label{fig3}}
\end{center}
\end{figure}

\begin{figure}
\begin{center}
\includegraphics[angle=0,width=0.65\textwidth]{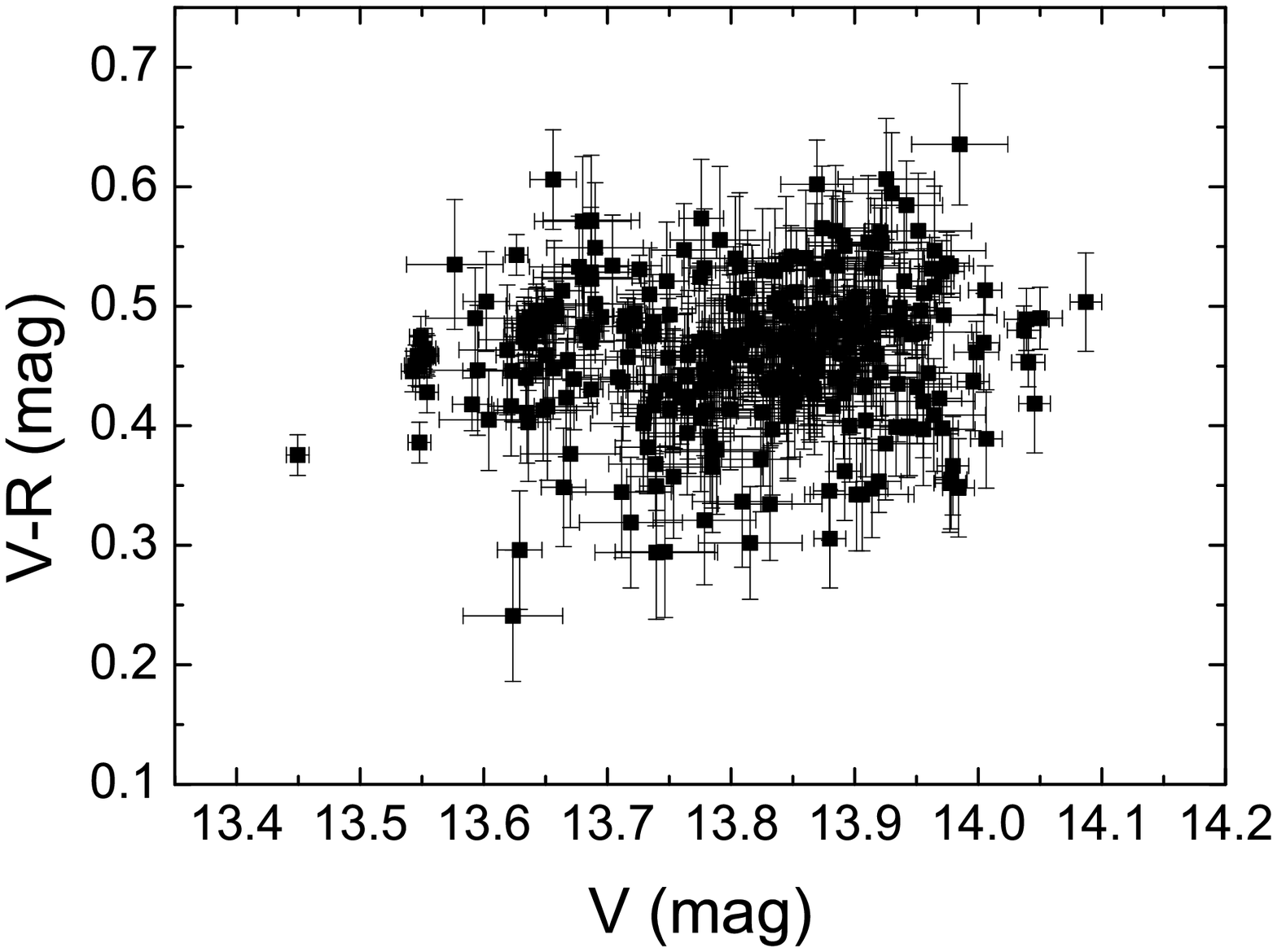}
\includegraphics[angle=0,width=0.65\textwidth]{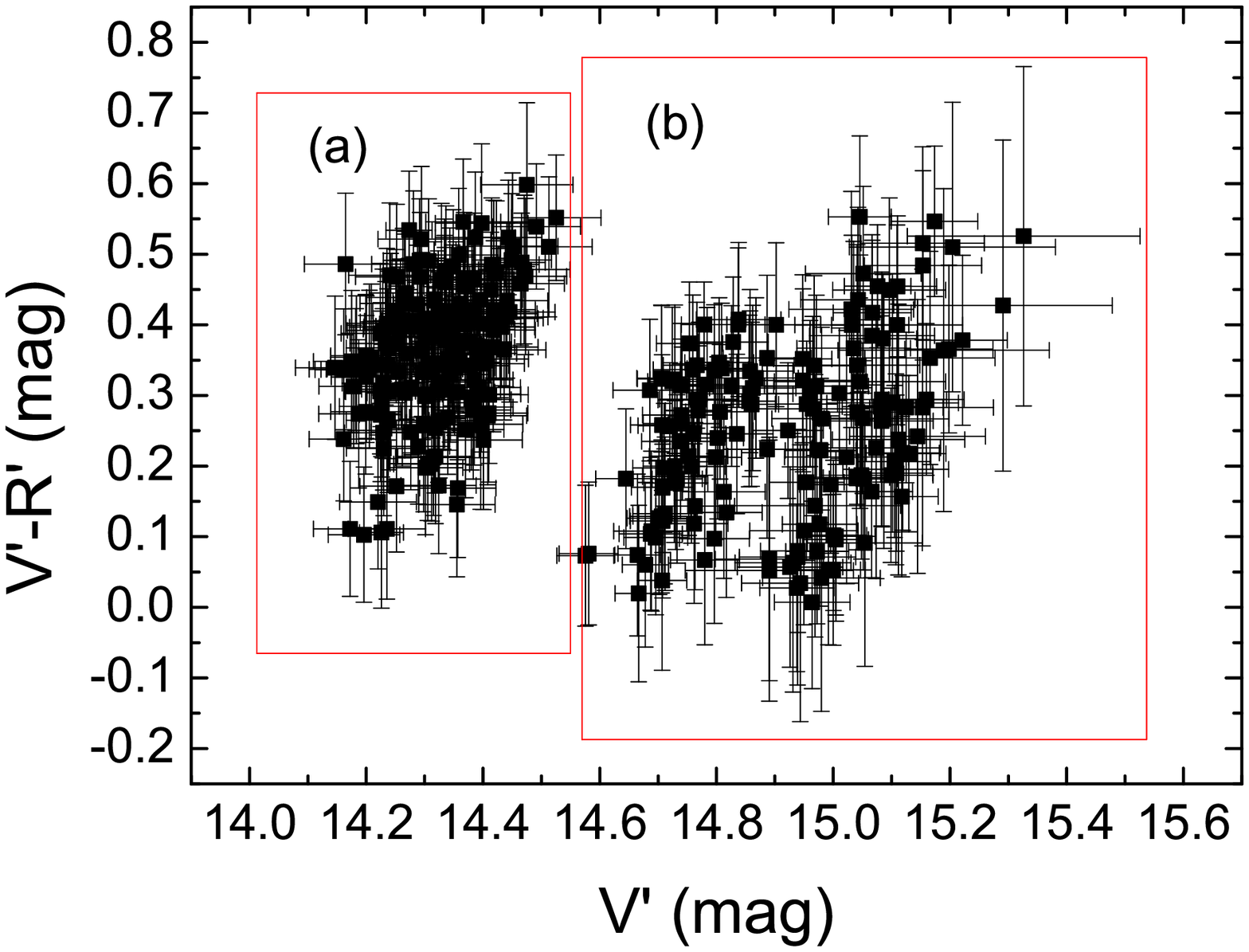}
\caption{The correlations between $V-R$ index and $V$ magnitude for
long-term timescale. The top panel is the results of non-correcting
the host galaxy contributions, and the bottom panel is the
correcting results.}
\end{center}
\end{figure}

\clearpage

\begin{deluxetable}{cccr}
\tablecaption{Observation log of photometric
observations\label{tbl-3}} \tablewidth{0pt}
\tablehead{\colhead{Date(UT)} & \colhead{Number(I,R,V)} &
\colhead{Time spans(h)} & \colhead{Time resolutions(min)}}
\startdata
2010 04 01  &   17,17,17    &   3.1 &   12  \\
2010 04 02  &   20,20,20    &   4.1 &   13  \\
2010 04 03  &   18,18,18    &   3.7 &   13  \\
2012 04 22  &   18,18,15    &   4.3 &   14  \\
2012 04 23  &   20,20,20    &   4.5 &   14  \\
2012 04 24  &   21,21,21    &   4.8 &   14  \\
2012 04 25  &   3,0,2   &   0.6 &   19  \\
2013 04 01  &   10,9,9  &   2.8 &   17  \\
2013 04 03  &   15,15,13    &   3.9 &   16  \\
2013 04 04  &   5,6,5   &   1   &   14  \\
2014 04 17  &   0,21,0  &   2   &   5   \\
2014 04 19  &   11,8,7  &   3.6 &   20  \\
2014 04 20  &   8,9,6   &   3.5 &   25  \\
2014 04 22  &   11,10,9 &   4.4 &   25  \\
2014 04 23  &   9,10,9  &   4.4 &   25  \\
2014 04 24  &   19,16,12    &   3.9 &   13  \\
2014 04 25  &   5,5,0   &   2.5 &   25  \\
2015 04 06  &   15,18,17    &   3.2 &   14  \\
2015 04 07  &   8,3,0   &   2.1 &   13  \\
2015 04 08  &   19,18,18    &   3.9 &   13  \\
2015 04 13  &   16,13,9 &   3.9 &   13  \\
2015 04 14  &   14,14,17    &   3.6 &   13  \\
2015 04 15  &   21,19,21    &   4.5 &   13  \\
2015 05 16  &   20,19,20    &   3.4 &   9   \\
2015 05 17  &   18,17,15    &   3.3 &   10  \\
2015 05 19  &   11,12,11    &   3.4 &   12  \\
2015 05 24  &   28,22,24    &   6.8 &   10  \\
2015 05 25  &   44,38,39    &   8   &   10  \\
\enddata
\end{deluxetable}

\begin{deluxetable}{ccccr}
\tablecaption{Data of $I$ Band\label{tbl-2}} \tablewidth{0pt}
\tablehead{\colhead{Date(UT)} & \colhead{MJD} & \colhead{Magnitude}
& \colhead{$\sigma$}& \colhead{$F_{host}$(mJy)}} \startdata
2010 Apr 01 &   55287.782   &   12.539  &   0.015   &   11.1    \\
2010 Apr 01 &   55287.790   &   12.521  &   0.015   &   12  \\
2010 Apr 01 &   55287.798   &   12.528  &   0.015   &   11.6    \\
2010 Apr 01 &   55287.806   &   12.466  &   0.015   &   12.4    \\
2010 Apr 01 &   55287.814   &   12.478  &   0.015   &   12.4    \\
2010 Apr 01 &   55287.822   &   12.472  &   0.015   &   12.4    \\
2010 Apr 01 &   55287.830   &   12.473  &   0.015   &   12.4    \\
2010 Apr 01 &   55287.838   &   12.473  &   0.015   &   12.4    \\
\enddata
\tablecomments{Column (1) is the universal time (UT) of observation,
column (2) the corresponding modified Julian day (MJD), column (3)
the magnitude, column (4) the rms error, column (5) the fluxes of
host galaxy in mJy. Table 2 is available in its entirety in the
electronic edition of the {\sl The Astrophysical Journal
Supplement}. A portion is shown here for guidance regarding its form
and content.}
\end{deluxetable}

\begin{deluxetable}{ccccr}
\tablecaption{Data of $R$ Band\label{tbl-3}} \tablewidth{0pt}
\tablehead{\colhead{Date(UT)} & \colhead{MJD} & \colhead{Magnitude}
& \colhead{$\sigma$}& \colhead{$F_{host}$(mJy)}} \startdata
2010 Apr 01 &   55287.779   &   13.125  &   0.014   &   12.4    \\
2010 Apr 01 &   55287.787   &   13.129  &   0.014   &   12.4    \\
2010 Apr 01 &   55287.795   &   13.140  &   0.014   &   12.4    \\
2010 Apr 01 &   55287.803   &   13.139  &   0.014   &   12.8    \\
2010 Apr 01 &   55287.811   &   13.136  &   0.014   &   12.8    \\
2010 Apr 01 &   55287.819   &   13.134  &   0.014   &   12.8    \\
2010 Apr 01 &   55287.827   &   13.135  &   0.014   &   12.8    \\
2010 Apr 01 &   55287.835   &   13.134  &   0.014   &   12.8    \\
\enddata
\tablecomments{The meanings of columns are same with Table 2. Table
3 is available in its entirety in the electronic edition of the {\sl
The Astrophysical Journal Supplement}. A portion is shown here for
guidance regarding its form and content.}
\end{deluxetable}

\begin{deluxetable}{ccccr}
\tablecaption{Data of $V$ Band\label{tbl-4}} \tablewidth{0pt}
\tablehead{\colhead{Date(UT)} & \colhead{MJD} & \colhead{Magnitude}
& \colhead{$\sigma$}& \colhead{$F_{host}$(mJy)}} \startdata
2010 Apr 01 &   55287.776   &   13.679  &   0.009   &   11.5    \\
2010 Apr 01 &   55287.784   &   13.604  &   0.009   &   12.4    \\
2010 Apr 01 &   55287.792   &   13.600  &   0.009   &   12  \\
2010 Apr 01 &   55287.800   &   13.600  &   0.009   &   12.8    \\
2010 Apr 01 &   55287.808   &   13.597  &   0.009   &   12.8    \\
2010 Apr 01 &   55287.816   &   13.602  &   0.009   &   12.8    \\
2010 Apr 01 &   55287.824   &   13.606  &   0.009   &   12.8    \\
2010 Apr 01 &   55287.832   &   13.604  &   0.009   &   12.8    \\
\enddata
\tablecomments{The meanings of columns are same with Table 2. Table
4 is available in its entirety in the electronic edition of the {\sl
The Astrophysical Journal Supplement}. A portion is shown here for
guidance regarding its form and content.}
\end{deluxetable}

\begin{deluxetable}{ccccccccccr}
\tablecaption{Results of IDV Observations of Mrk 501\label{tbl-5}}
\tablewidth{0pt} \tablehead{\colhead{Date} & \colhead{Band} &
\colhead{N}  & \colhead{$C$} & \colhead{$F$} &\colhead{$F_C(99)$}
&\colhead{$F_A$} &\colhead{$F_A(99)$} &\colhead{V/N} &\colhead{A(\%)} &\colhead{Ave(mag)}\\
\colhead{(1)} & \colhead{(2)} & \colhead{(3)}  & \colhead{(4)} &
\colhead{(5)} &\colhead{(6)} &\colhead{(7)} &\colhead{(8)}
&\colhead{(9)} &\colhead{(10)} &\colhead{(11)}} \startdata
2010 Apr 01 &   I   &   17  &   3.74    &   14.02   &   3.37    &   2.92    &   5.41    &   PV  &   21.62   &   12.47   \\
2010 Apr 01 &   R   &   17  &   1.50    &   2.26    &   3.37    &   1.62    &   5.41    &   N   &   8.60    &   13.14   \\
2010 Apr 01 &   V   &   17  &   4.25    &   18.10   &   3.37    &   1.81    &   5.41    &   PV  &   17.65   &   13.60   \\
2010 Apr 02 &   I   &   20  &   1.98    &   3.90    &   3.03    &   4.99    &   4.70    &   PV  &   7.34    &   12.54   \\
2010 Apr 02 &   R   &   20  &   1.75    &   3.07    &   3.03    &   1.02    &   4.70    &   PV  &   12.01   &   13.20   \\
2010 Apr 02 &   V   &   20  &   1.45    &   2.10    &   3.03    &   1.21    &   4.70    &   PV  &   4.46    &   13.70   \\
2010 Apr 03 &   I   &   18  &   4.07    &   16.61   &   3.24    &   9.67    &   5.06    &   V   &   10.94   &   12.63   \\
2010 Apr 03 &   R   &   18  &   4.00    &   16.03   &   3.24    &   2.01    &   5.06    &   PV  &   11.21   &   13.26   \\
2010 Apr 03 &   V   &   18  &   3.64    &   13.24   &   3.24    &   15.02   &   5.06    &   V   &   13.69   &   13.76   \\
2012 Apr 22 &   I   &   18  &   1.05    &   1.12    &   3.24    &   2.06    &   5.06    &   N   &   8.33    &   12.70   \\
2012 Apr 22 &   R   &   18  &   1.76    &   3.17    &   3.24    &   2.16    &   5.06    &   N   &   14.69   &   13.35   \\
2012 Apr 22 &   V   &   15  &   0.97    &   0.95    &   3.70    &   0.53    &   5.99    &   N   &   5.50    &   13.82   \\
2012 Apr 23 &   I   &   20  &   3.02    &   9.22    &   3.03    &   11.37   &   4.70    &   V   &   24.59   &   12.84   \\
2012 Apr 23 &   R   &   20  &   3.82    &   14.63   &   3.03    &   12.38   &   4.70    &   V   &   25.76   &   13.48   \\
2012 Apr 23 &   V   &   20  &   3.83    &   14.66   &   3.03    &   14.65   &   4.70    &   V   &   20.59   &   13.97   \\
2012 Apr 24 &   I   &   21  &   5.83    &   33.98   &   2.94    &   11.21   &   4.46    &   V   &   30.91   &   12.78   \\
2012 Apr 24 &   R   &   21  &   3.88    &   15.03   &   2.94    &   15.99   &   4.46    &   V   &   25.46   &   13.48   \\
2012 Apr 24 &   V   &   21  &   4.96    &   24.62   &   2.94    &   8.47    &   4.46    &   V   &   30.04   &   13.97   \\
2013 Apr 01 &   I   &   10  &   1.31    &   1.76    &   5.35    &   1.75    &   9.55    &   N   &   13.68   &   12.77   \\
2013 Apr 01 &   R   &   9   &   1.65    &   2.80    &   6.03    &   0.98    &   10.93   &   N   &   13.20   &   13.42   \\
2013 Apr 01 &   V   &   9   &   1.30    &   1.84    &   6.03    &   0.67    &   10.93   &   N   &   16.50   &   13.91   \\
2013 Apr 03 &   I   &   15  &   2.47    &   6.09    &   3.70    &   2.01    &   5.99    &   PV  &   15.95   &   12.82   \\
2013 Apr 03 &   R   &   15  &   2.58    &   6.72    &   3.70    &   3.70    &   3.70    &   PV  &   18.43   &   13.41   \\
2013 Apr 03 &   V   &   13  &   1.59    &   2.53    &   4.16    &   1.57    &   6.99    &   N   &   14.27   &   13.94   \\
2014 Apr 17 &   R   &   21  &   1.05    &   1.11    &   2.94    &   0.51    &   4.46    &   N   &   12.72   &   13.44   \\
2014 Apr 19 &   I   &   11  &   1.77    &   3.14    &   2.94    &   8.94    &   8.65    &   PV  &   14.10   &   12.82   \\
2014 Apr 20 &   R   &   9   &   1.33    &   1.77    &   6.03    &   7.78    &   10.93   &   N   &   11.96   &   13.41   \\
2014 Apr 22 &   I   &   11  &   1.54    &   2.39    &   4.85    &   19.38   &   8.65    &   PV  &   9.39    &   12.82   \\
2014 Apr 22 &   R   &   10  &   1.02    &   1.14    &   5.35    &   1.85    &   9.55    &   N   &   10.93   &   13.46   \\
2014 Apr 23 &   I   &   9   &   3.51    &   12.38   &   6.03    &   8.47    &   10.93   &   PV  &   25.41   &   12.77   \\
2014 Apr 23 &   R   &   10  &   3.45    &   11.93   &   5.35    &   12.20   &   9.55    &   V   &   20.92   &   13.40   \\
2014 Apr 23 &   V   &   9   &   1.92    &   3.79    &   6.03    &   37.94   &   10.93   &   PV  &   18.84   &   13.90   \\
2014 Apr 24 &   I   &   19  &   2.61    &   6.84    &   3.13    &   27.54   &   4.86    &   V   &   25.20   &   12.75   \\
2014 Apr 24 &   R   &   16  &   1.79    &   3.29    &   3.52    &   9.30    &   5.67    &   PV  &   22.69   &   13.37   \\
2014 Apr 24 &   V   &   12  &   2.16    &   4.70    &   4.46    &   9.66    &   7.59    &   PV  &   28.29   &   13.85   \\
2015 Apr 06 &   I   &   15  &   1.19    &   1.45    &   3.70    &   5.97    &   5.99    &   N   &   12.63   &   12.61   \\
2015 Apr 08 &   I   &   19  &   1.22    &   1.52    &   3.13    &   8.05    &   4.86    &   PV  &   12.17   &   12.67   \\
2015 Apr 08 &   R   &   18  &   1.25    &   1.58    &   3.24    &   1.33    &   5.06    &   N   &   11.27   &   13.42   \\
2015 Apr 08 &   V   &   18  &   1.41    &   2.00    &   3.24    &   1.54    &   5.06    &   N   &   21.56   &   13.93   \\
2015 Apr 13 &   I   &   16  &   1.19    &   1.44    &   3.52    &   4.14    &   5.67    &   N   &   18.11   &   12.53   \\
2015 Apr 13 &   R   &   13  &   1.37    &   1.88    &   4.16    &   0.76    &   6.99    &   N   &   16.29   &   13.19   \\
2015 Apr 13 &   V   &   9   &   1.23    &   1.65    &   6.03    &   8.04    &   10.93   &   N   &   9.92    &   13.70   \\
2015 Apr 14 &   I   &   14  &   1.64    &   2.73    &   3.91    &   1.96    &   6.55    &   N   &   10.62   &   12.63   \\
2015 Apr 14 &   R   &   14  &   0.96    &   0.99    &   3.91    &   1.68    &   6.55    &   N   &   11.31   &   13.31   \\
2015 Apr 14 &   V   &   17  &   2.69    &   7.22    &   3.37    &   4.90    &   5.40    &   PV  &   14.73   &   13.76   \\
2015 Apr 15 &   I   &   21  &   3.94    &   15.58   &   2.94    &   29.75   &   4.46    &   V   &   24.84   &   12.90   \\
2015 Apr 15 &   R   &   19  &   4.78    &   22.81   &   3.13    &   5.34    &   4.86    &   V   &   37.95   &   13.51   \\
2015 Apr 15 &   V   &   21  &   2.05    &   4.27    &   2.94    &   3.15    &   4.46    &   PV  &   30.58   &   13.92   \\
2015 May 16 &   I   &   20  &   1.50    &   2.26    &   3.03    &   1.49    &   4.70    &   N   &   18.10   &   12.99   \\
2015 May 16 &   R   &   19  &   1.19    &   1.43    &   3.13    &   2.38    &   4.86    &   N   &   12.21   &   13.63   \\
2015 May 16 &   V   &   20  &   4.32    &   18.71   &   3.03    &   0.31    &   4.70    &   PV  &   20.67   &   14.01   \\
2015 May 17 &   I   &   18  &   2.05    &   4.19    &   3.24    &   4.88    &   5.06    &   PV  &   25.90   &   12.84   \\
2015 May 17 &   R   &   17  &   1.38    &   1.93    &   3.37    &   2.33    &   5.41    &   N   &   14.54   &   13.45   \\
2015 May 17 &   V   &   15  &   2.18    &   4.81    &   3.70    &   3.18    &   5.99    &   PV  &   32.61   &   13.87   \\
2015 May 19 &   I   &   11  &   2.48    &   6.18    &   4.85    &   1.70    &   8.65    &   PV  &   16.89   &   12.80   \\
2015 May 19 &   R   &   12  &   1.19    &   1.45    &   4.46    &   0.96    &   7.59    &   N   &   10.74   &   13.43   \\
2015 May 19 &   V   &   11  &   1.35    &   1.86    &   4.85    &   0.45    &   8.65    &   N   &   11.65   &   13.91   \\
2015 May 24 &   I   &   28  &   1.60    &   2.60    &   2.51    &   3.54    &   3.63    &   PV  &   14.95   &   12.84   \\
2015 May 24 &   R   &   22  &   1.33    &   1.77    &   2.86    &   4.41    &   4.32    &   PV  &   16.26   &   13.41   \\
2015 May 24 &   V   &   24  &   0.99    &   1.02    &   2.72    &   2.06    &   4.03    &   N   &   13.39   &   13.88   \\
2015 May 25 &   I   &   44  &   1.40    &   2.01    &   2.06    &   8.92    &   2.79    &   PV  &   20.79   &   12.77   \\
2015 May 25 &   R   &   38  &   0.97    &   1.02    &   2.20    &   1.39    &   3.02    &   N   &   12.83   &   13.38   \\
2015 May 25 &   V   &   39  &   1.00    &   1.06    &   2.16    &   3.25    &   2.96    &   PV  &   12.33   &   13.85   \\

\enddata
\tablecomments{Column 1 is the date of observation, column 2 the
observed band, column 3 the number of data points, column 4 the
value of $C$ test, column 5 the average $F$ value, column 6 the
critical $F$ value with 99 per cent confidence level, column 7 the
$F$ value of ANOVA, column 8 the critical $F$ value of ANOVA with 99
per cent confidence level, column 9 the variability status (V:
variable, PV: probable variable, N: non-variable), column 10-11 the
variability amplitude and daily average magnitudes respectively.}
\end{deluxetable}

\begin{deluxetable}{cccccr}
\tablecaption{Results of Magnitude Change \label{tbl-4}}
\tablewidth{0pt} \tablehead{\colhead{Date(UT)} & \colhead{MJD} &
\colhead{$\bigtriangleup$T(min)} & \colhead{Band} &
\colhead{$\bigtriangleup$ (Var)}& \colhead{$\bigtriangleup^{'}$
(Var)}} \startdata
2015 Apr 15 &  $57127.743-57127.914$ & 246 & $I$ & 0.25 (V)   &   0.29 (V)  \\
            &  $57127.884-57127.920$ & 52 & $R$ & 0.32 (V)   &   0.18 (V)  \\
            &  $57127.875-57127.884$ & 13 & $R$ & 0.21 (V)   &   0.09 (V)  \\
2014 Apr 24 &  $56771.753-56771.851$ & 141 & $I$ & 0.26 (V)   &   0.20 (N)  \\
2014 Apr 23 &  $56770.726-56770.781$ & 79 & $R$ & 0.15 (V)   &   0.12 (N)  \\
2012 Apr 24 &  $56041.710-56041.750$ & 58 & $I$ & 0.22 (V)   &   0.12 (V)  \\
            &  $56041.717-56041.747$ & 43 & $R$ & 0.14 (V)   &   0.03 (V)  \\
            &  $56041.707-56041.757$ & 72 & $R$ & 0.09 (V)   &   0.31 (V)  \\
            &  $56041.882-56041.892$ & 14 & $V$ & 0.18 (V)   &   0.09 (N)  \\
2012 Apr 23 &  $56040.736-56040.914$ & 256 & $I$ & 0.25 (V)   &   0.11 (N)  \\
            &  $56040.763-56040.773$ & 14 & $R$ & 0.11 (V)   &   0.13 (V)  \\
            &  $56040.730-56040.740$ & 14 & $V$ & 0.13 (V)   &   0.19 (V)  \\
2010 Apr 03 &  $55289.787-55289.850$ & 91 & $I$ & 0.11 (V)   &   0.1 (N)  \\
            &  $55289.763-55289.898$ & 194 & $V$ & 0.14 (V)   &   0.03 (V)  \\
\enddata
\tablecomments{$\bigtriangleup$ and $\bigtriangleup^{'}$ are
Magnitude change for non-correcting and correcting Galactic
extinction and the host galaxy contributions. V/N stands for whether
the night is intraday variability or not.}
\end{deluxetable}

\begin{deluxetable}{ccccr}
\tablecaption{Results of Linear Regression Analysis. $r$ and $r'$
are the coefficient of correlation for non-correcting and correcting
the host galaxy contributions; $P$ and $P'$ are the chance
probability for non-correcting and correcting the host galaxy
contributions.\label{tbl-4}} \tablewidth{0pt}
\tablehead{\colhead{Date(UT)} & \colhead{$r$} & \colhead{$P$} &
\colhead{$r'$} & \colhead{$P'$}} \startdata
2010 Apr 01 & 0.82 & $<0.0001$ &0.34  &0.2  \\
2010 Apr 02 & 0.92 & $<0.0001$ & 0.52&0.02 \\
2010 Apr 03 & 0.57 & 0.01 &0.77 &0.0002 \\
2010 Apr 01-03 & 0.64 & $<0.0001$ & 0.61 & $<0.0001$ \\
2012 Apr 22 & 0.68 & 0.01 &0.72 &0.002 \\
2012 Apr 24 & 0.48 & 0.03 & 0.32&0.16 \\
2012 Apr 22-24 &0.46 &0.0004 &0.62 &$<0.0001$ \\
2013 Apr 03 & 0.42 & 0.2 &0.61 & 0.03 \\
2013 Apr 01-04 &0.65 &0.0003 &0.71 &$<0.0001$ \\
2014 Apr 24 & 0.55 & 0.08 & 0.72&0.01 \\
2014 Apr 19-May 08 & 0.27 & 0.08 & 0.51&0.0003 \\
2015 Apr 06 & 0.54 & 0.03 &0.57 & 0.02\\
2015 Apr 08 & 0.82 & $<0.0001$ & 0.65& 0.005\\
2015 Apr 14 & 0.87 & 0.0001 & 0.78&0.002 \\
2015 Apr 15 & 0.57 & 0.02 & 0.63 & 0.007 \\
2015 May 16 & 0.75 & 0.0006 &0.74 &0.0006 \\
2015 May 17 & 0.83 & 0.0001 & 0.6&0.02 \\
2015 May 24 & 0.85 & $<0.0001$ &0.81 &0.0003 \\
2015 May 25 & 0.78 & $<0.0001$ & 0.82& $<0.0001$\\
2015 Apr 06-May 25 & -0.01 & 0.89 & 0.31& 0.0002\\
\enddata
\end{deluxetable}

\end{document}